\documentclass[a4paper,11pt]{article}
\pdfoutput=1 

\usepackage{jinstpub} 
\usepackage{siunitx}
\usepackage{amsmath}
\usepackage{verbatim}
\usepackage{overpic}
\usepackage{url}
\usepackage{xurl}
\usepackage{xspace}
\usepackage{subfigure}
\usepackage[symbol]{footmisc}

\usepackage{listings}
\usepackage{xcolor}

\definecolor{bgcolor}{rgb}{0.96,0.96,0.94}
\definecolor{modulegreen}{rgb}{0.0,0.5,0.0}
\definecolor{commentblue}{rgb}{0.0,0.2,0.7}
\definecolor{stringred}{rgb}{0.7,0.1,0.1}

\lstdefinestyle{allpix}{
  backgroundcolor=\color{bgcolor},
  basicstyle=\ttfamily\small,
  commentstyle=\color{commentblue}\itshape,
  stringstyle=\color{stringred},
  frame=single,
  breaklines=true,
  showstringspaces=false,
  numbers=right,
  numberstyle=\tiny\color{gray},
  numbersep=8pt
}
\usepackage{hyperref}
\hypersetup{colorlinks,breaklinks,
  linkcolor=black,citecolor=blue,
  filecolor=blue,urlcolor=blue,
  pdfpagemode=UseNone
}

\usepackage{caption}

\usepackage{chngcntr}
\counterwithout{equation}{section}

\graphicspath{{./}{figure/}} 

\newcommand*{\tab}[1]{Table~\ref{tab:#1}}
\newcommand*{\fig}[1]{Figure~\ref{fig:#1}}
\newcommand*{\sect}[1]{Section~\ref{sect:#1}}
\newcommand*{\app}[1]{Appendix~\ref{app:#1}}

\newcommand*{\lst}[1]{Listing~\ref{lst:#1}}

\usepackage{lineno}

\title{\boldmath TCAD + $\text{Allpix}^2$ simulation study of MALTA2 on Czochralski substrate, a Depleted Monolithic Active Pixel Sensor for future tracking}

\author[a,*]{L. Li,\note{Corresponding author.}}
\author[a]{K. Krizka, }
\author[c]{P. Behera, }
\author[d]{D.V. Berlea, }
\author[e]{D. Bortoletto, }
\author[f]{C. Buttar, }
\author[c]{T. Chembakan, }
\author[g]{V. Dao, }
\author[c]{G. Dash, }
\author[h]{Y. Enari, }
\author[d]{L. Fasselt, }
\author[b]{S. Haberl, }
\author[b]{T. Inada, }
\author[i]{F.K. Isik, }
\author[d]{C. Issever, }
\author[j]{X. Li, }
\author[h]{Y. Okazaki, }
\author[b]{H. Pernegger, }
\author[b]{P. Riedler, }
\author[b]{W. Snoeys, }
\author[b]{C.A Solans Sanchez, }
\author[b]{A. Swoboda, }
\author[i]{I. Turk Cakir, }
\author[b]{M. van Rijnbach, }
\author[c]{A. Vijay, }
\author[d]{S. Worm}

\affiliation[a]{University of Birmingham, Edgbaston Park Rd, B15 2TT, Birmingham, United Kingdom}
\affiliation[b]{CERN, Esplanade des Particules 1, 1211, Meyrin, Switzerland}
\affiliation[c]{Indian Institute Technology Madras, Hostel Ave, Tamil Nadu 600036, Chennai, India}
\affiliation[d]{Deutsches Elektronen-Synchrotron DESY, Platanenallee 6, 15738, Zeuthen, Germany}
\affiliation[e]{University of Oxford, Keble Road, OX1 3RH, Oxford, United Kingdom}
\affiliation[f]{University of Glasgow, University Ave, G12 8QQ, Glasgow, United Kingdom}
\affiliation[g]{Stony Brook University, 100 Nicolls Road, Stony Brook, NY 11794, New York, United States of America}
\affiliation[h]{High Energy Accelerator Research Organization, 1-1 Oho, Tsukuba, Ibaraki, 305-0801, Japan}
\affiliation[i]{Ankara University, Ankara, Turkey}
\affiliation[j]{Los Alamos National Laboratory, United States of America}

\emailAdd{long.l@cern.ch}

\abstract{In this work, a hybrid simulation framework combining TCAD and $\text{Allpix}^2$ is applied to investigate 
the sensor performance of MALTA2 on Czochralski substrate, a depleted monolithic active pixel sensor designed for future 
tracking. The study begins with 3D modeling and transient simulation in TCAD, with 
generic doping profiles and simple well structures. The user-defined doping profiles and simulated electric fields then 
serve as inputs for high-statistics Monte Carlo simulations in $\text{Allpix}^2$.

Simulations show that the sensor performance, particularly detection efficiency, cluster size and time response, depends 
on the doping concentration of the shallow low‑dose N‑type implant. The doping concentration is optimized by comparing 
simulated results with measurement data. The active depth of the depleted region of the MALTA2 sensor is estimated 
in both simulations and measurements using a grazing angle method, in which the sensor is positioned at various 
inclinations relative to the beam, covering angles from $0^{\circ}$ to $60^{\circ}$. At the optimal doping concentration, the 
simulated active depth agrees with measurements to within 4\,$\upmu$m. 

Consequently, this study demonstrates that MALTA2 performance can be studied using generic doping profiles, 
without requiring proprietary foundry information.}


\arxivnumber{2605.23860} 

\begin{document}
\maketitle
\flushbottom

\section{Introduction}
\label{sect:intro}
MALTA2 \cite{MALTA2_FE} is the second-generation prototype sensor of the MALTA series, which is a Depleted Monolithic Active Pixel Sensor 
(DMAPS) designed for precise tracking in high hit-rate environment of future collider experiments. It is fabricated in a Tower 180\,nm 
CMOS image sensor (CIS) technology on either a high resistivity epitaxial layer (EPI, > 1\,k$\Omega \cdot$cm) grown on a substrate or a 
high resistivity Czochralski substrate (Cz, 3-4\,k$\Omega \cdot$cm), 
featuring a matrix of $224\times512$ pixels with a pitch of 36.4\,\text{$\upmu$}m $\times$ 36.4\,\text{$\upmu$}m \cite{MALTA2_FE,MALTA2_radhard}. 
Sensor thickness ranges from 50 to 300\,$\upmu$m on EPI substrate and 100 to 300\,$\upmu$m on Cz, depending on the variant.
A shallow low-dose N-doped blanket (STD) was introduced 
to achieve large depletion region \cite{MALTA_process_modification}. Additionally, two process modifications were implemented for better 
charge collection: a gap in the N-blanket (NGAP) at the pixel edge or an extra P-type implant at the same location (XDPW) \cite{MALTA_XDPW_NGAP}.

A grazing angle study \cite{GrazingAngleTechnique} was performed on MALTA2 using the CERN SPS mixed hadron beam, in which the samples were in 
Cz substrate with thickness of 100\,$\upmu$m and XDPW modification \cite{MALTA2Cz_grazing}. The detection efficiency and cluster size were 
measured as a function of the grazing angle, on both unirradiated and irradiated samples at a fluence of 
$1\,\times\,10^{15}\,1\,\text{MeV}\,\text{n}_{\text{eq}}/\text{cm}^2$ for non-ionising energy loss. Accordingly, the active depth of the 
depleted region was estimated.

Comprehensive simulation frameworks \cite{TCAD+APSQ_guideline} play an indispensable role in the R\&D cycle of advanced CMOS sensors 
such as MALTA2. First, 
compared with multiple foundry shuttles or large-scale test-beam campaigns, TCAD + $\text{Allpix}^2$ simulation offers a highly 
cost-effective and time-efficient alternative, as they eliminate the recurring costs of wafer fabrication and beam-time allocation 
during the initial parameter screening phase. Second, the predictive capability of this approach allows 
designers to explore the physical mechanisms underlying charge collection and signal formation before any hardware is manufactured. 
Furthermore, the simulation environment offers unparalleled flexibility for 
systematic parameter scans, such as doping profiles, which are often difficult or prohibitively time-consuming to cover experimentally. 
Motivated by these advantages, the performance of MALTA2 on high-resistivity Cz substrate with XDPW modification is studied using a
generic-doping-profile-based TCAD + $\text{Allpix}^2$ simulation framework. To validate the simulation and optimize the input parameters, 
the simulated sensor properties, in particular detection efficiency, cluster size and time response, are compared with test-beam measurements.

The manuscript is organised as follows: a brief introduction of the workflow is given in \sect{framework}; 3D modeling and 
transient simulations in TCAD are elaborated in \sect{tcad}, $\text{Allpix}^2$ simulations together 
with comparisons with test-beam data are demonstrated in \sect{allpix}; and a conclusion is given in \sect{conclusion}.

\section{TCAD + $\text{Allpix}^2$ simulation framework for MALTA2}
\label{sect:framework}

\begin{figure}[htb!]
  \centering
  \includegraphics[width=0.7\textwidth, origin=c]{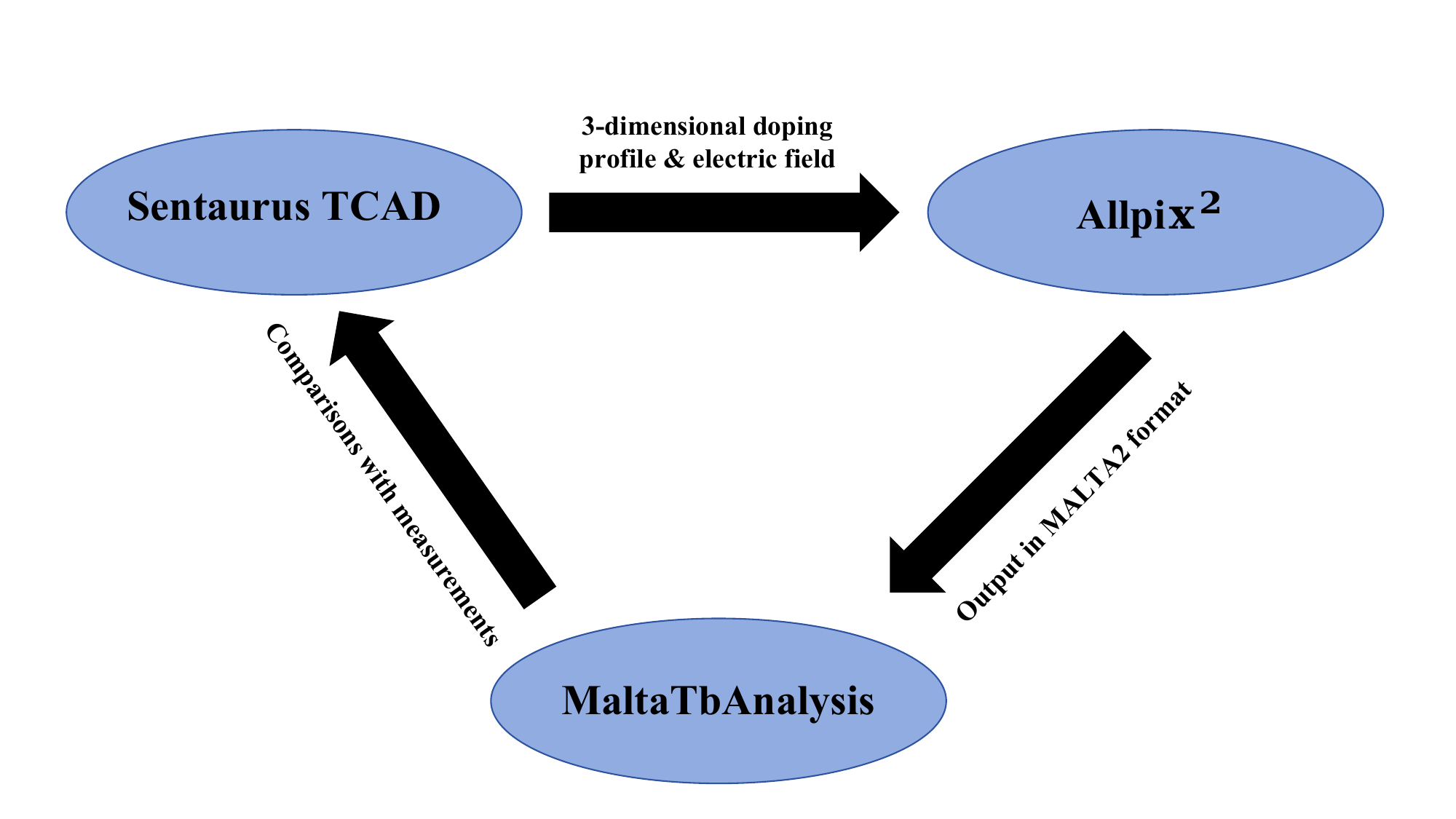}

  \caption{Workflow of the TCAD + $\text{Allpix}^2$ simulation framework for MALTA2.}
  \label{fig:workflow}
\end{figure}

The workflow of the TCAD + $\text{Allpix}^2$ simulation framework for MALTA2 is illustrated in \fig{workflow}. 
Sentaurus Technology Computer Aided Design (TCAD) software from Synopsys\cite{TCAD} is used for device modeling and transient simulation to 
evaluate the electrical behavior of the MALTA2 sensor. First, the 3D structures of MALTA (for telescope reference planes) \cite{MALTA} and 
MALTA2 devices are modelled in TCAD. Next, transient simulations are performed on 
these device models. Finally, the relevant data, including the user-defined doping profiles and the electric field distributions obtained from 
the simulation results, are formatted and stored as \texttt{DF-ISE} files.

In $\text{Allpix}^2$, the doping profiles and electric field distributions are read from the \texttt{DF-ISE} files and converted into mesh models 
to define the physical configurations of the sensor. High-statistics Monte Carlo simulations are then performed for both device under test (DUT) only 
and full-telescope configurations, as detailed in \sect{allpix}.

In these simulations, the \texttt{DepositionGeant4} module is used to simulate the energy deposition of 120\,GeV/c protons in the sensor, 
via an interface to Geant4 \cite{G4_1, G4_2, G4_3}. The \texttt{GenericPropagation} module then handles the propagation of 
charge carriers through the sensor, based on the imported electric field and doping profiles. For digitization, the \texttt{CSADigitizer} module 
is employed as a behavioral proxy of the MALTA2 voltage-sensitive amplifier. Additionally, in the full-telescope configuration, a custom module, 
\texttt{Malta2TreeWriter}, is developed to format the spatial and temporal hit information following the MALTA readout scheme \cite{MALTA_readout} 
for compatibility with the analysis chain. Details of the modules and configurations used in this work can be found in \app{appendix}.

In \texttt{MaltaTbAnalysis}, the reconstruction and analysis software for MALTA2, both test-beam data and full-telescope simulation outputs are processed. 
The performance metrics, including detection efficiency, cluster size and time response, are calculated and recorded. The user-defined parameters, 
in particular the doping concentrations, are then iteratively tuned based on the comparisons between measurements and simulations, such that the simulated 
performance approaches the observed behaviors.

\section{Transient simulations in TCAD}
\label{sect:tcad}

\subsection{The sensor structures and assumptions}

\label{sect:tcad:geo}
\begin{figure}[htb!]
  \centering
  \includegraphics[width=0.6\textwidth, origin=c]{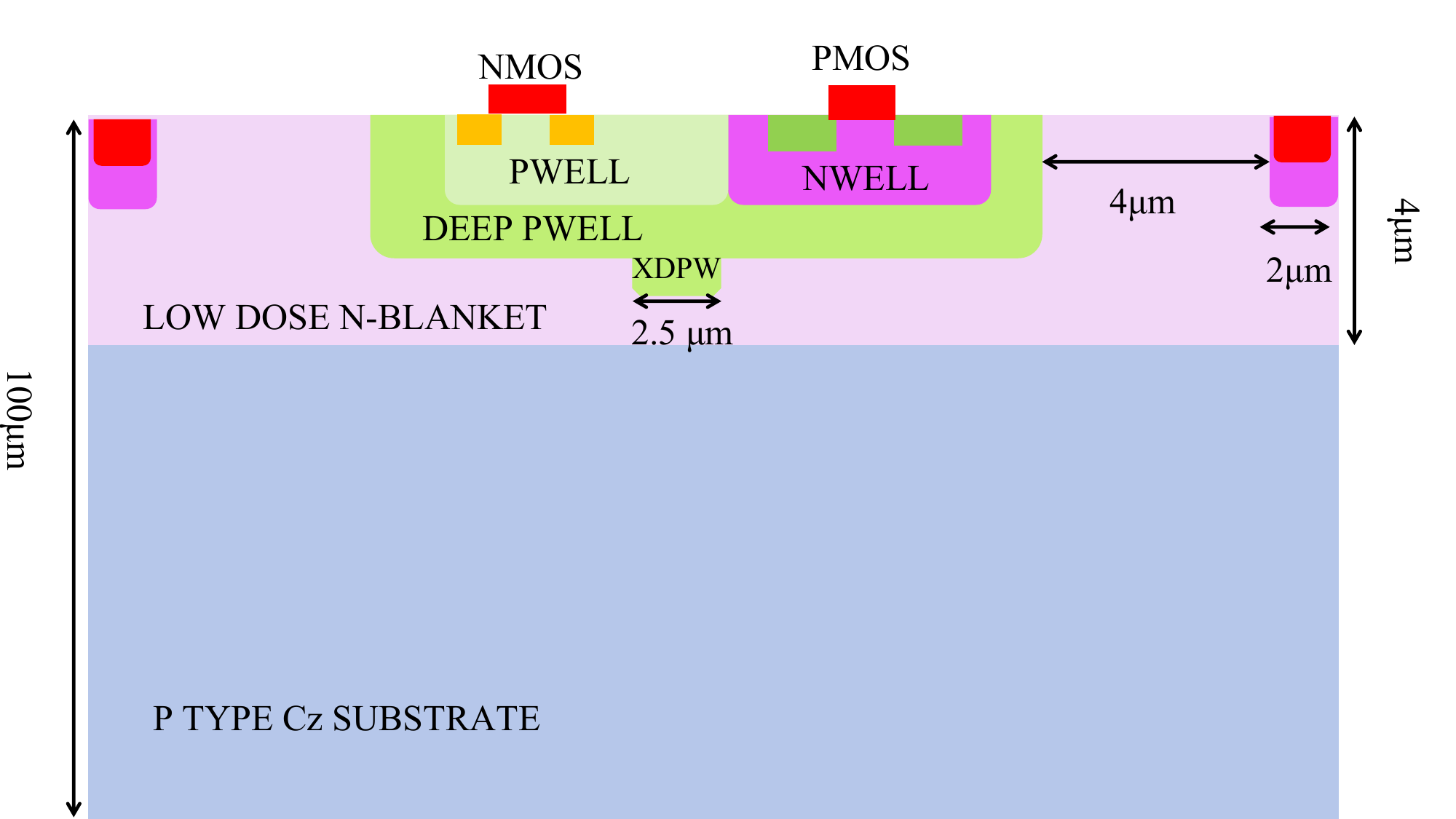}

  \caption{Cross section of the simulated sensor structure with Cz substrate and XDPW modification (not to scale). 
  The MOS transistors are not implemented in Sentaurus TCAD.}
  \label{fig:crossX}
\end{figure}

The TCAD simulation begins with the 3D device modeling in \texttt{Sentaurus Structure Editor} (SDE), where the sensor
structure is defined by the shapes and materials. The materials used in the simulation are: aluminum for the
electrodes, Si$\text{O}_2$ for the dielectrics and silicon for the sensor bulk. A silicon cuboid with dimension 
of $36.4 \times 36.4 \times 100\,\upmu\text{m}^3$ is defined as the sensor substrate. A uniform P-type doping  
concentration of $4.423 \times 10^{12}\,\text{cm}^{-3}$, corresponding to a substrate resistivity of approximately
3\,k$\Omega\cdot$cm, is assumed throughout the bulk. As shown in \fig{crossX}, a uniformly N-type doped layer 
(N-blanket) with a thickness of 4 $\upmu$m is introduced on top of the substrate. At the interface between the
substrate and the N-blanket, an error-function doping profile is applied along the depth direction to emulate 
dopant diffusion over a depth of 0.5 $\upmu$m. Within the N-blanket, an octagonal N-well diode with an equivalent 
diameter of 2\,$\upmu$m is embedded to serve as the charge-collecting node. In the cross-section, it appears as a
rectangular N-well region. A deep P-well (DPW), shown as a green rectangle, is also formed within the N-blanket, 
occupying the majority of its area, with a 4 $\upmu$m spacing maintained between the DPW and the N-well diode to 
ensure proper isolation of the collecting node. An extra deep P-well with a width of 2.5~$\upmu$m is added at 
the pixel edge, under the DPW. It should be noted that the MOS transistors and their associated 
wells are omitted in our model; therefore, no readout electronics are included.

Doping concentration plays a critical role in pixel sensor design, as it governs the key electrical properties of 
the device, particularly the electric field distribution within the sensor. However, the exact doping values are 
not provided by the foundry. In the absence of proprietary process details, typical doping concentrations are 
chosen with reference to the methodology described in Ref. \cite{TCAD+APSQ_guideline}, which presents a 
comprehensive simulation 
approach for a different project targeting the same technology \cite{MALTA2_FE,TJ_Investigator}. In our TCAD models, 
the doping concentrations for 
the wells range from $10^{15}\,\text{cm}^{-3}$ to $10^{19}\,\text{cm}^{-3}$, following a multi-Gaussian
superimposed distribution in the depth direction. In the lateral direction, an error-function profile is used 
to define the well boundaries. The N-blanket is uniformly doped, with the doping concentration 
(referred to as N-dop) inside the range from $1\,\times\,10^{14} \text{cm}^{-3}$ to $5\,\times10^{15} \text{cm}^{-3}$.
To investigate the impact of N-dop on sensor performance, six concentration values are selected for comparison, 
defined as A, B (= 1.5\,A), C (= 2.0\,A), D (= 2.5\,A), E (= 3.0\,A), and F (= 4.0\,A). Concentration  
B is used for demonstration in the following unless otherwise specified.

Once the geometry and doping profiles are defined, the refinement parameters are set for \texttt{Sentaurus Mesh} (SMesh). 
A finer mesh improves simulation accuracy, but the simulation time increases with the mesh size. In this work, 
two types of meshing strategies are used: a global refinement with a fixed element size of 
5\,$\times$ 5\,$\times$ 5\,$\upmu\text{m}^3$ applied to the whole device, and a doping-gradient-dependent refinement 
with a maximum size of 0.5\,$\times$ 0.5\,$\times$ 0.5\,$\upmu\text{m}^3$ and a minimum size of 
0.05\,$\times$ 0.05\,$\times$ 0.05\,$\upmu\text{m}^3$ applied in wells and interfaces. The total number of elements 
is constrained to 1 million, considering the CPU and memory costs.

\subsection{Transient simulations with heavy ion model}
\label{sect:tcad:transient}
The device models were simulated in both quasi-stationary and transient modes using \textit{Sentaurus Device} (SDevice).
In the quasi-stationary simulation, the DC operating point was determined and the electric field and carrier distributions 
were calculated after ramping the bias voltage of p-well and substrate to -6\,V (-30\,V on substrate for telescope planes 3 and 4). 
Then, the responses of collection electrodes to a charged particle traversing 
the device were simulated based on the standard drift--diffusion transport, doping--dependent mobility 
(Masetti model \cite{MasettiMobility}) and Shockley--Read--Hall recombination \cite{Shockley, HALL}, in transient mode. 
A \textit{HeavyIon} model was applied to emulate the penetration of the charged particle. The particle was set to
traverse normally from the back-side plane with a path length of 100\,$\upmu$m and a transverse impact parameter of 
1\,$\upmu$m. The linear energy transfer factor ($\text{LET}_\text{f}$) was set to $9.6 \times 10^{-6}$ pC/$\upmu$m, 
corresponding to the energy deposition of minimum ionising particles (MIPs) in silicon \cite{PDG}.  

\begin{figure}[htb!]
  \centering
  \subfigure[]{
    \includegraphics[width=0.28\textwidth]{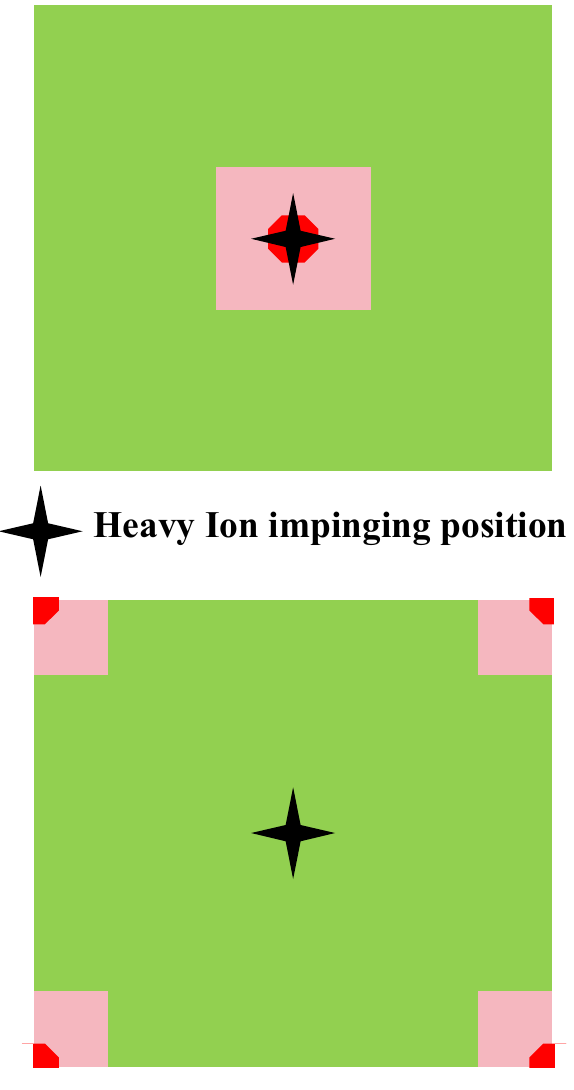}
    \label{fig:hi:layout}
  }
  \subfigure[]{
    \includegraphics[width=0.68\textwidth]{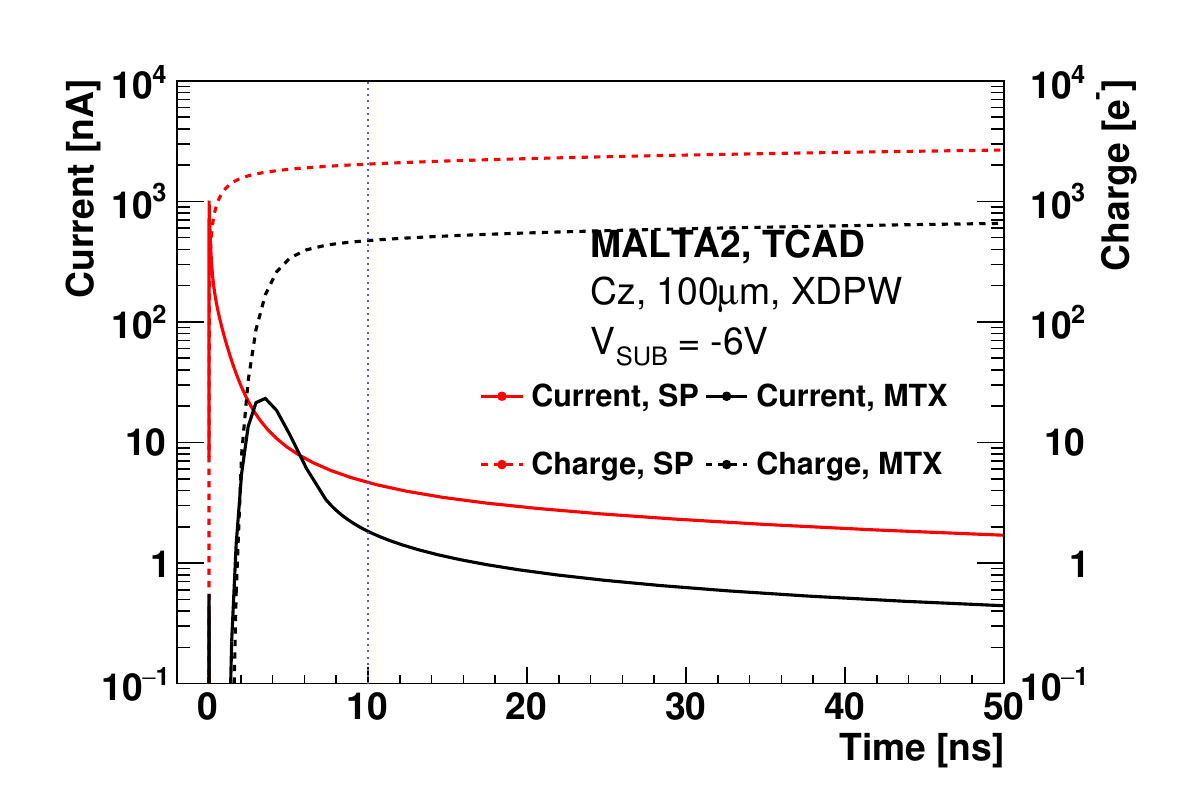}
    \label{fig:hi:transient}
  }
  \caption{(a) Layouts of a single pixel (top) and a $2\,\times\,2$ pixel matrix (bottom). The red octagon represents 
  the N-well collection diode, the green region indicates the DPW, and the pink area denotes the gap 
  between the N-well and the DPW. In the matrix layout, only one quarter of the N-well diode is implemented in 
  each pixel. The black star marks the impact point of the heavy ion. (b) Transient current responses (solid lines) and 
  corresponding charge collections (dashed lines) to the heavy-ion penetration for matrix (black) and single pixel 
  (red) layouts. The bias voltage is -6\,V. Since the responses of all pixels in the matrix are identical, 
  only the response of one pixel is shown.}
  \label{fig:hi}
\end{figure}

As shown in \fig{hi:layout},  two pixel layouts are designed for two typical particle impingement cases: a single pixel 
(SP) with a hit placed on the collection diode, and a $2\,\times\,2$ pixel matrix (MTX) with a hit centered among the 
four pixels. The transient currents per pixel as a function of time for both scenarios are shown in \fig{hi:transient}, 
along with their time integrals, i.e., the collected charge. 
In the SP configuration, a pulse is generated immediately after particle penetration, with a peak amplitude reaching 1\,$\upmu$A.
The high amplitude and short pulse duration (FWHM < 0.1\,ns) indicate that the charge is collected almost 
instantaneously, as the electron-hole pairs generated beneath the N-well diode. In the MTX configuration, by contrast, a lower peak 
amplitude of approximately 20\,nA is observed, and the pulse peak is delayed by about 3.5\,ns relative to the SP case. 
Moreover, the pulse is significantly broader (FWHM > 2.2\,ns), suggesting a longer charge collection process 
as carriers drift from the impact point at the pixel corner toward the surrounding collection electrodes.
Regarding the charge collection for SP and MTX cases, approximately 2665\,$\text{e}^-$ and 2628\,$\text{e}^-$ 
(about 657\,$\text{e}^-$ per pixel in MTX configuration) are collected within a 50\,ns simulation window, respectively. 
More than 70\% of the charge is collected within the first 10\,ns in both scenarios. 
Consequently, the integration time in $\text{Allpix}^2$ is set to 10\,ns, based on the time response observed in 
transient simulations and compatibility with the integration time expected in MALTA2 design.

\section{Monte Carlo simulations in $\text{Allpix}^2$}
\label{sect:allpix}
Following the workflow described in \sect{framework}, the user-defined doping profiles and electric field distributions 
obtained from the TCAD simulations were imported into $\text{Allpix}^2$ for high-statistics Monte Carlo studies, including 
the analog study in DUT-only configuration and the tracking-based study in full-telescope configuration. 
Two sets of doping profiles and electric field distributions 
were used in $\text{Allpix}^2$: one derived from the SP layout, and the other from the MTX layout. Discrepancies 
in the simulated performance between the two scenarios were found to be negligible; therefore, 
the SP layout is adopted for all subsequent simulations. The operating bias voltage and threshold are set to -6\,V 
and 650\,$\text{e}^-$, respectively, unless otherwise specified. 

\begin{figure}[htb!]
  \centering
    \subfigure[]{
    \includegraphics[width=0.45\textwidth, origin=c]{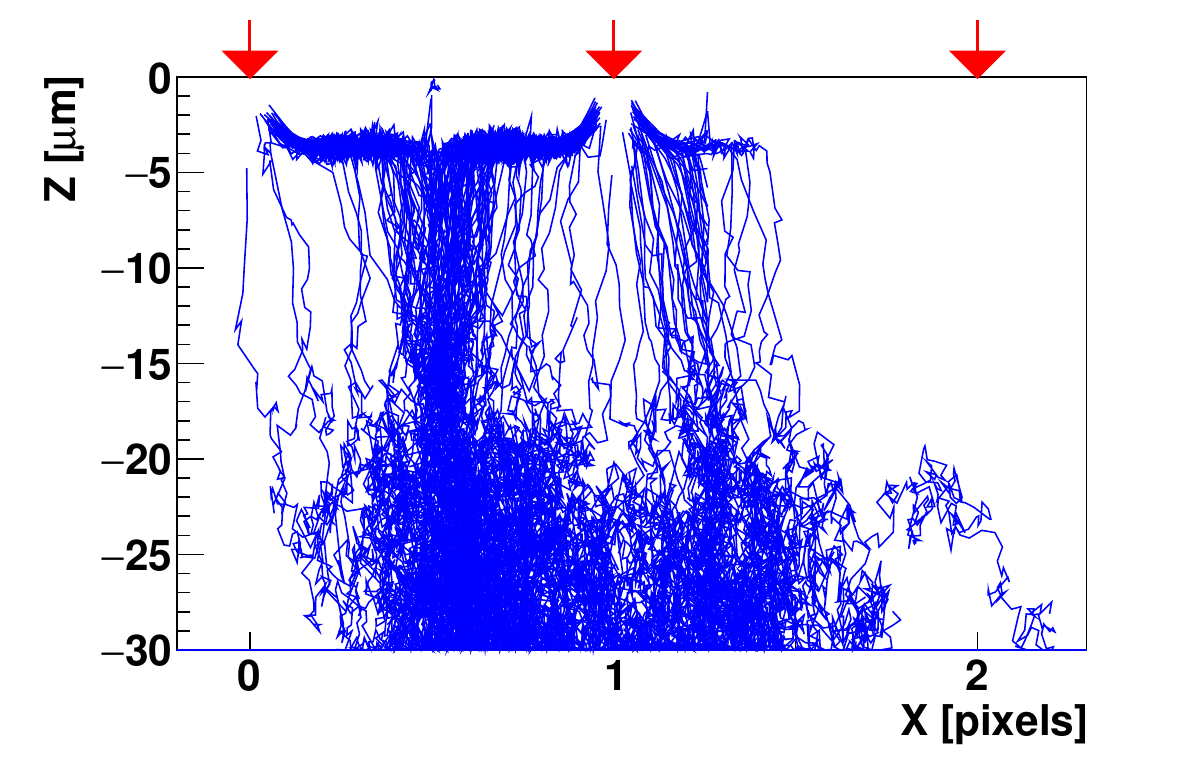}
    \label{fig:efield:linegraph}
  }%
  \subfigure[]{
    \includegraphics[width=0.45\textwidth]{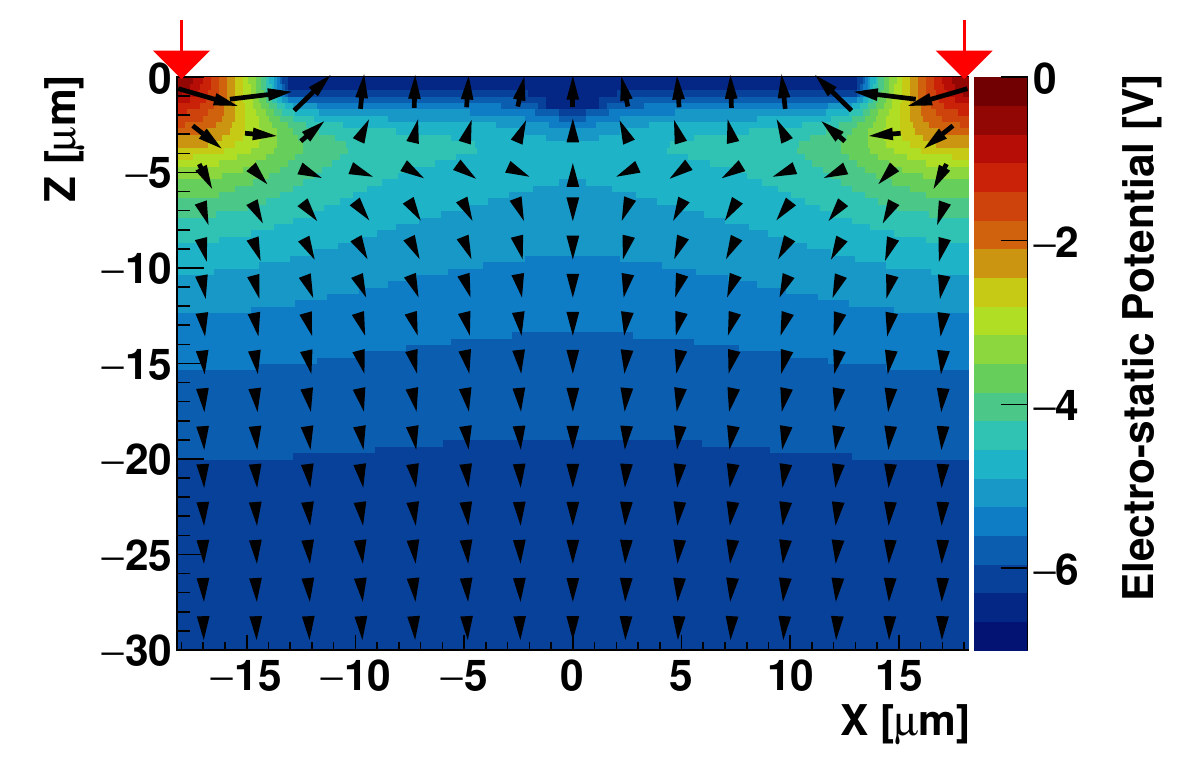}
    \label{fig:efield:distribution}
  }
  \caption{(a) Simulated electron trajectories for a single 120\,GeV/c proton event, based on the user-defined doping profile 
  and the resulting electric field after TCAD simulation. The image shows the region within the first 30\,$\upmu$m depth 
  from the sensor surface. The proton traverses the sensor along 
  Z‑direction, generating electron–hole pairs along its track (holes are omitted for clarity). 
  (b) Electrostatic potential distribution of the MALTA2 sensor from the TCAD simulation, also shown for the first 
  30\,$\upmu$m depth. The red arrows in both panels indicate the position of the N‑well collection diode, while the black 
  arrows in (b) represent the electric field vectors.
  }
  \label{fig:efield}
\end{figure}

\fig{efield:linegraph} shows an example of electron collection for a single 120\,GeV/c proton event in $\text{Allpix}^2$,
within the first 30\,$\upmu$m depth from the sensor surface. Following energy deposition by the traversing proton, 
the generated electrons drift vertically under the electric field shown in \fig{efield:distribution}, and are subsequently split 
laterally due to the enhanced horizontal electric field at the N-blanket/substrate interface and collected by the surrounding N-well
diode. This charge collection behavior is consistent with the electric field distribution obtained from the TCAD
simulation.

The MALTA series utilizes an asynchronous readout scheme to process digital hits \cite{MALTA_readout}. Pixels are grouped into $2 \times 8$ 
sub-arrays, and hits (pixels with collected charge above the threshold) within the same group are encoded into a single digital word, 
referred to as a MALTA word. All words arriving within a 1.6\,ns time window are merged via a bit-wise OR operation, which may result 
in hit displacement or loss \cite{MALTA2_FastSim}. Consequently, a hits-merging algorithm is included in this work.

\subsection{Simulations with DUT only}
\label{sect:allpix:fastsim}
DUT-only configured simulations were performed to study the N-dop dependent performance of MALTA2. 
In this configuration, a MALTA2 pixel matrix described in \sect{intro} is defined, 
with doping profiles and electric field distributions imported from the TCAD simulation for each of the six doping 
scenarios (A–F). For each simulated proton event, the analog outputs of a pixel, such as the collected charge
and time-of-arrival (ToA), are recorded if the collected charge exceeds the threshold. 

\begin{figure}[htb!]
  \centering
  \subfigure[]{
    \includegraphics[width=0.33\textwidth]{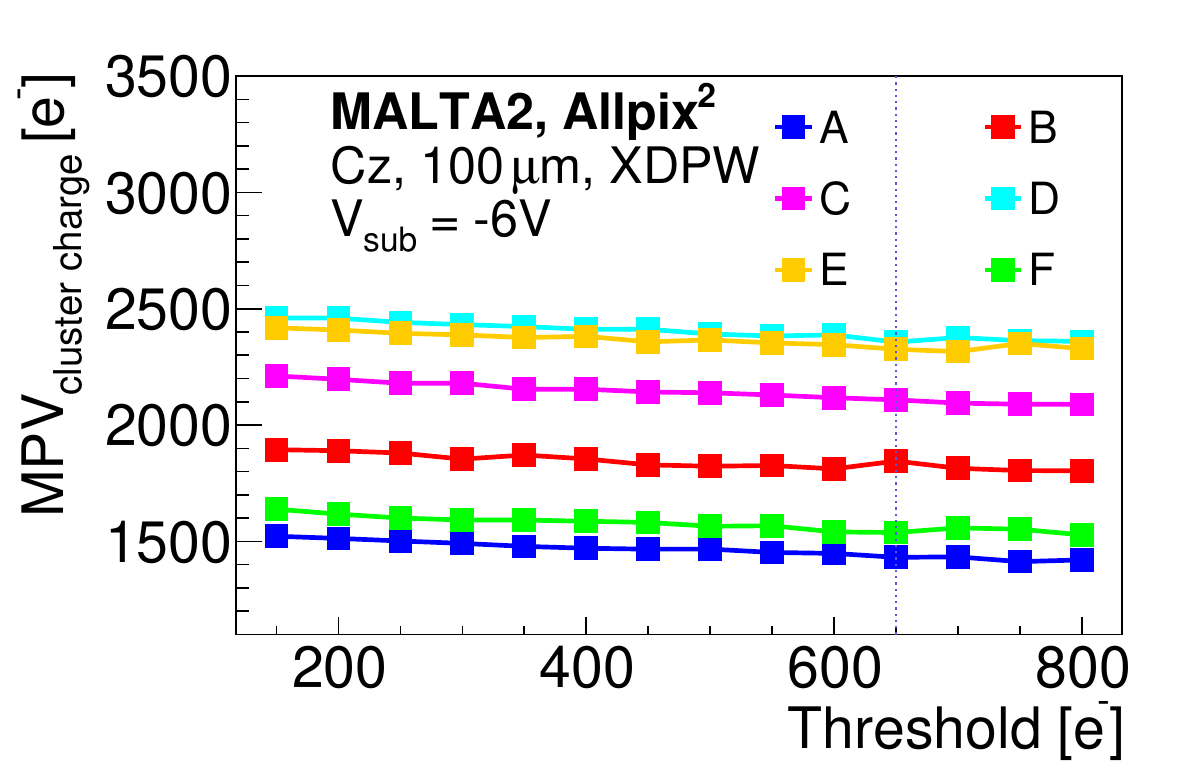}
    \label{fig:dut:npix}
  }%
  \subfigure[]{
    \includegraphics[width=0.33\textwidth]{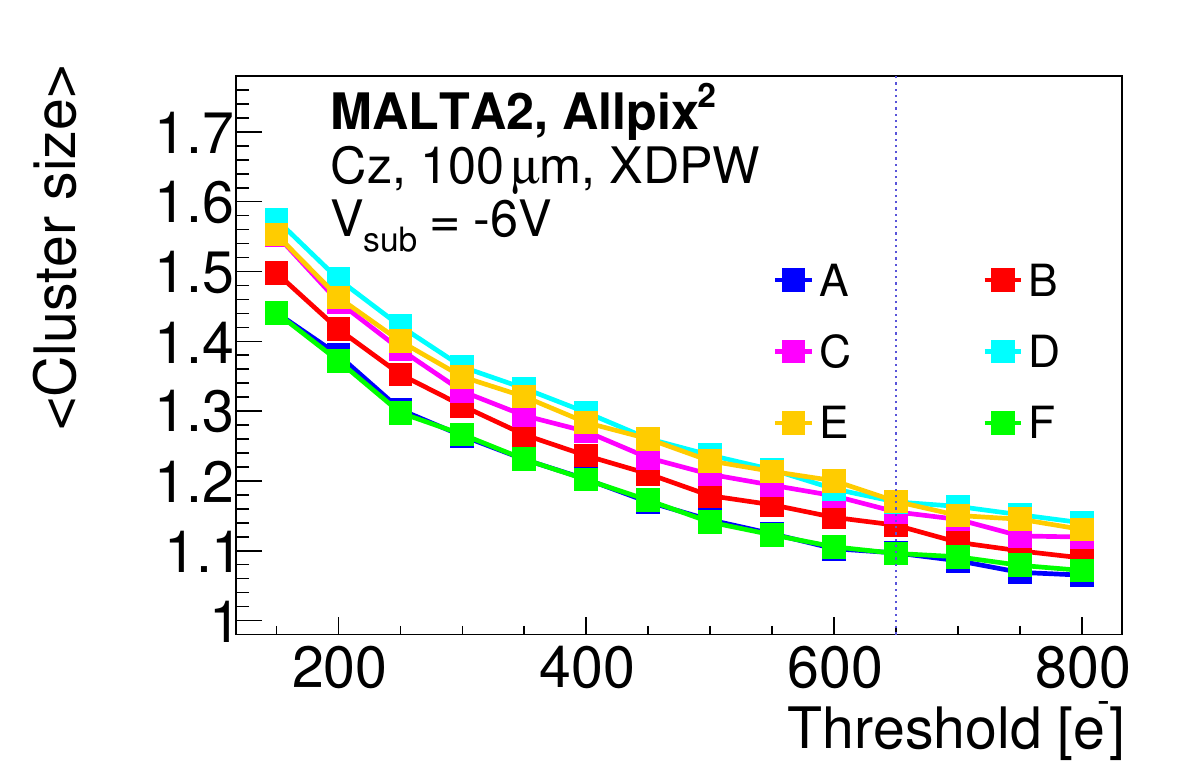}
    \label{fig:dut:clcharge}
  }%
  \subfigure[]{
    \includegraphics[width=0.33\textwidth]{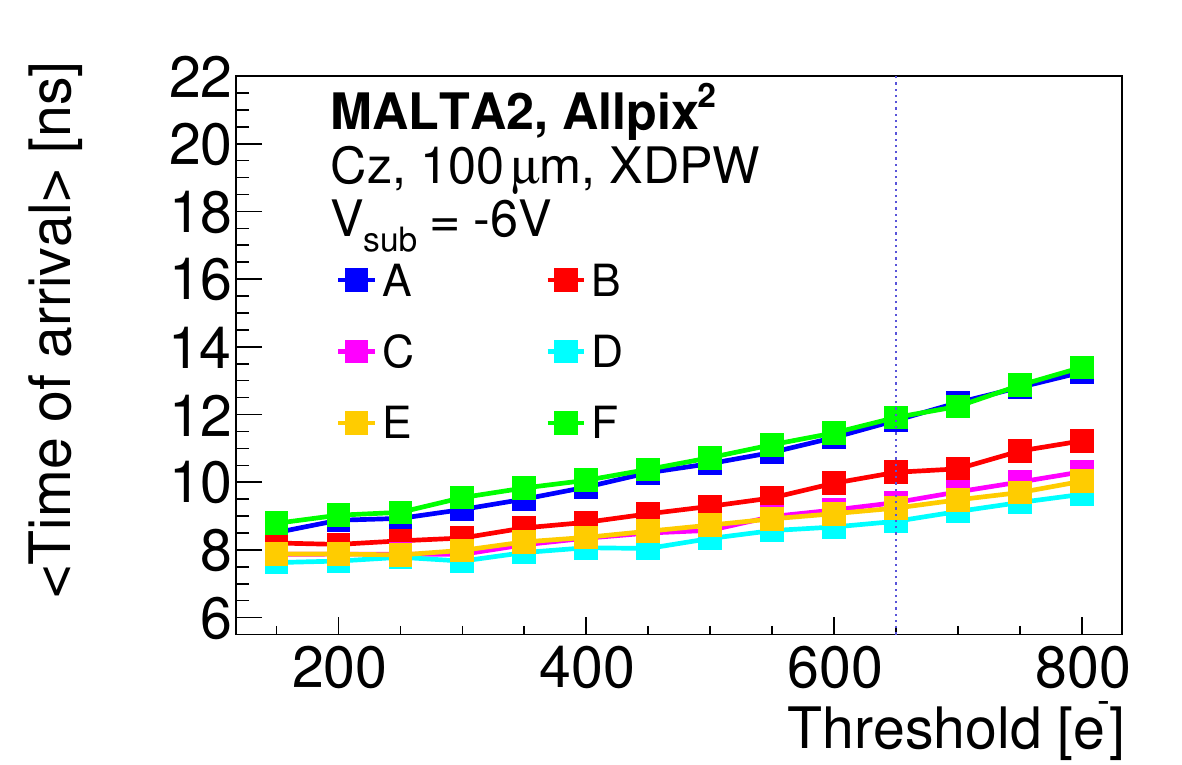}
    \label{fig:dut:pixtoa}
  }
  \subfigure[]{
    \includegraphics[width=0.33\textwidth]{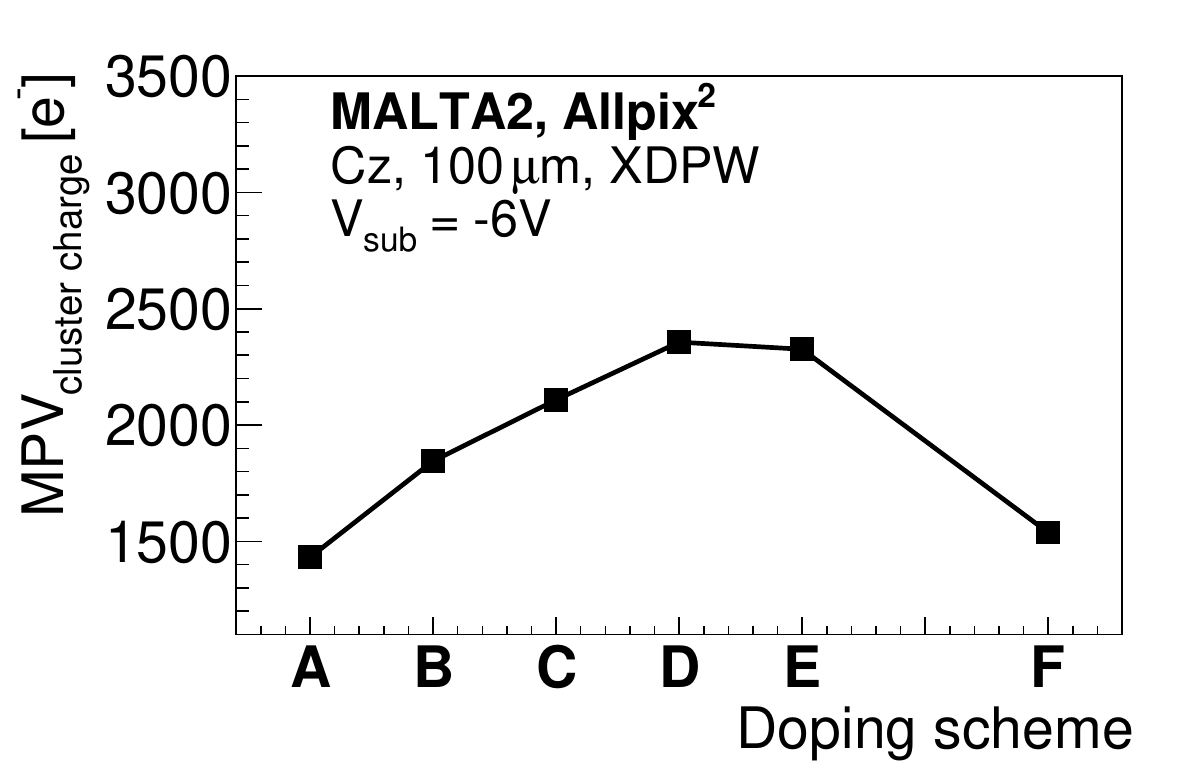}
    \label{fig:dut:npix_dop}
  }%
  \subfigure[]{
    \includegraphics[width=0.33\textwidth]{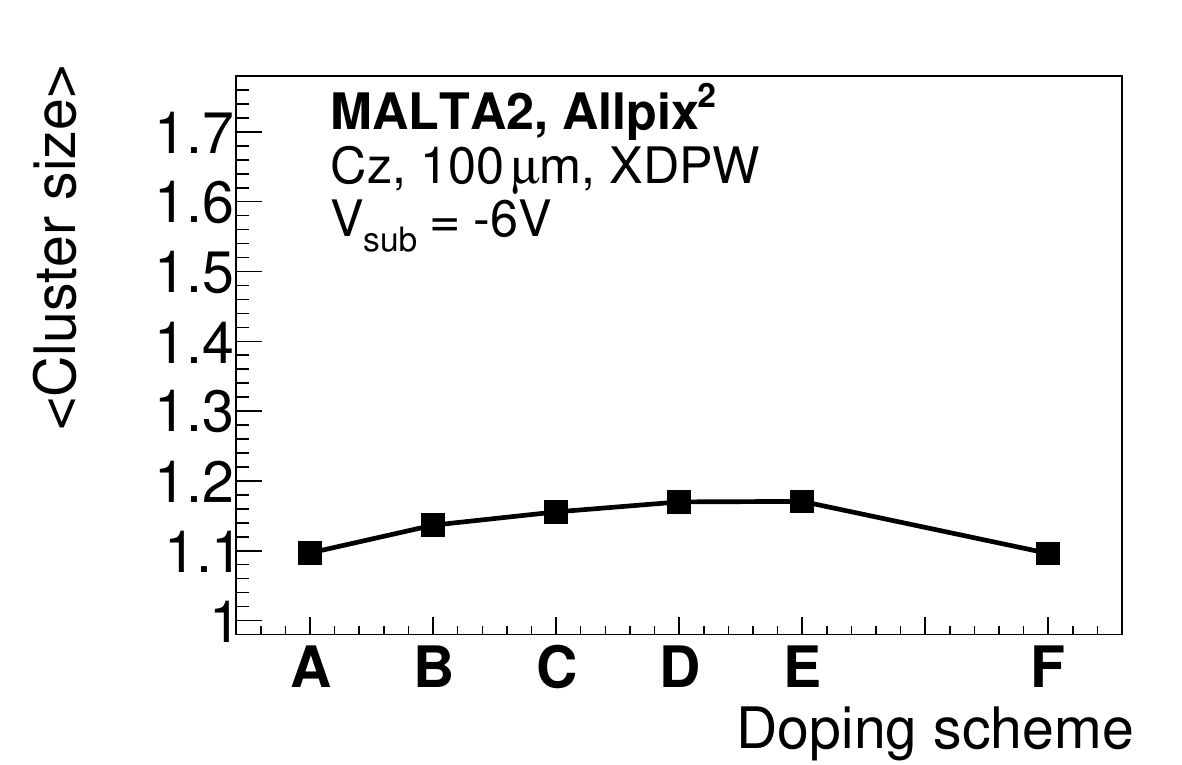}
    \label{fig:dut:clcharge_dop}
  }%
    \subfigure[]{
    \includegraphics[width=0.33\textwidth]{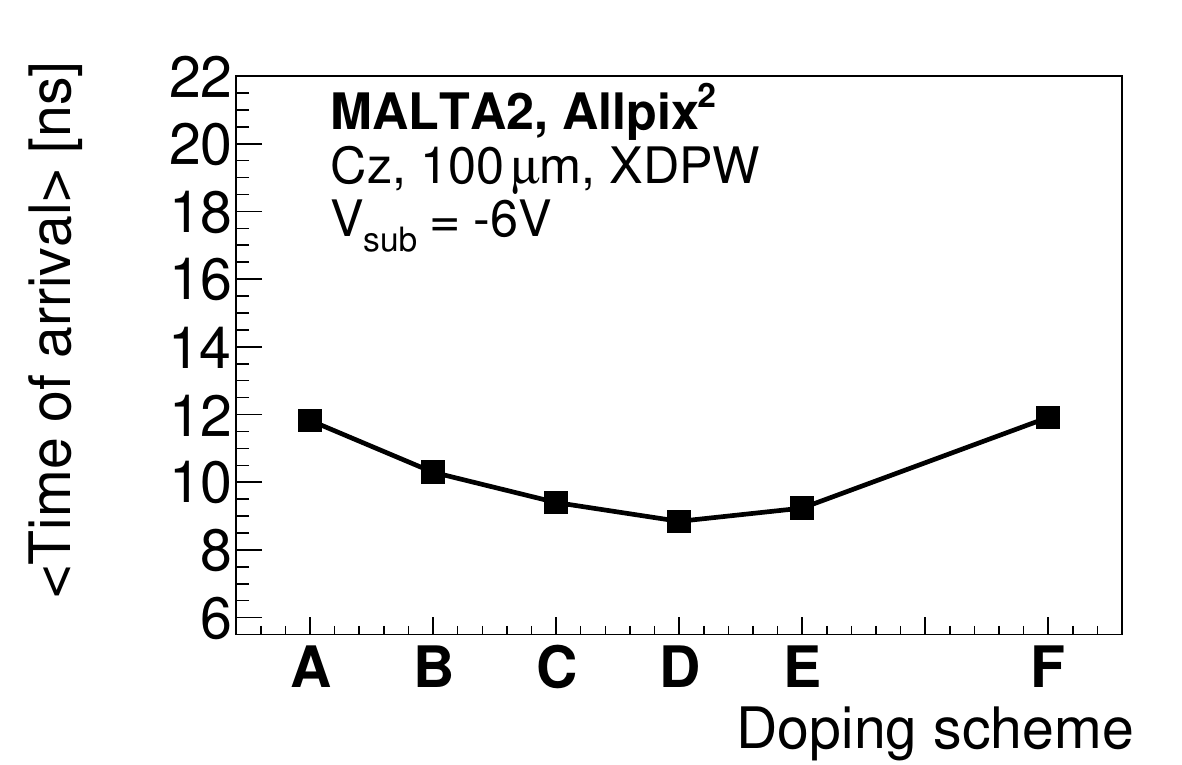}
    \label{fig:dut:pixtoa_dop}
  }%

  \caption{(a) the MPV of cluster charge, (b) the average cluster size and (c) the average hit 
  ToA, as a function of threshold for N-dop concentrations A to F. And (d) the 
  MPV of cluster charge, (e) the average cluster size, and (f) the average hit ToA as a 
  function of N-dop concentration, simulated at a fixed threshold of 650\,$\text{e}^-$. 
  The bias voltage is set to -6\,V.}
  \label{fig:dut}
\end{figure}

\fig{dut} (a) -- (c) show the most probable value (MPV) of cluster charge, the average cluster size 
and the average hit ToA, as a function of the threshold for N-dop concentrations A through F.  
For a given N-dop, as expected, the MPV of cluster charge and the averge cluster size decreases whereas the average ToA 
increases with increasing threshold. 

Variations of these properties with respect to the N-dop concentration are more intriguing. The MPV
of cluster charge, the average cluster size and the average hit ToA, as a function of the N-dop, 
at a threshold of 650\,$\text{e}^-$, are shown in \fig{dut} (d) -- (f). 
In the range of A-D, the MPV of cluster charge and the average cluster size increase as the N-dop rises, indicating enhanced charge
collection and sharing. Concurrently, the decrease in average ToA with increasing N-dop indicates a faster time response. However, the
charge collection and timing performance degrade as the N-dop increases further, showing a turning point at N-dop D.

\begin{figure}[htb!]
  \centering
  \subfigure[]{
    \includegraphics[width=0.48\textwidth]{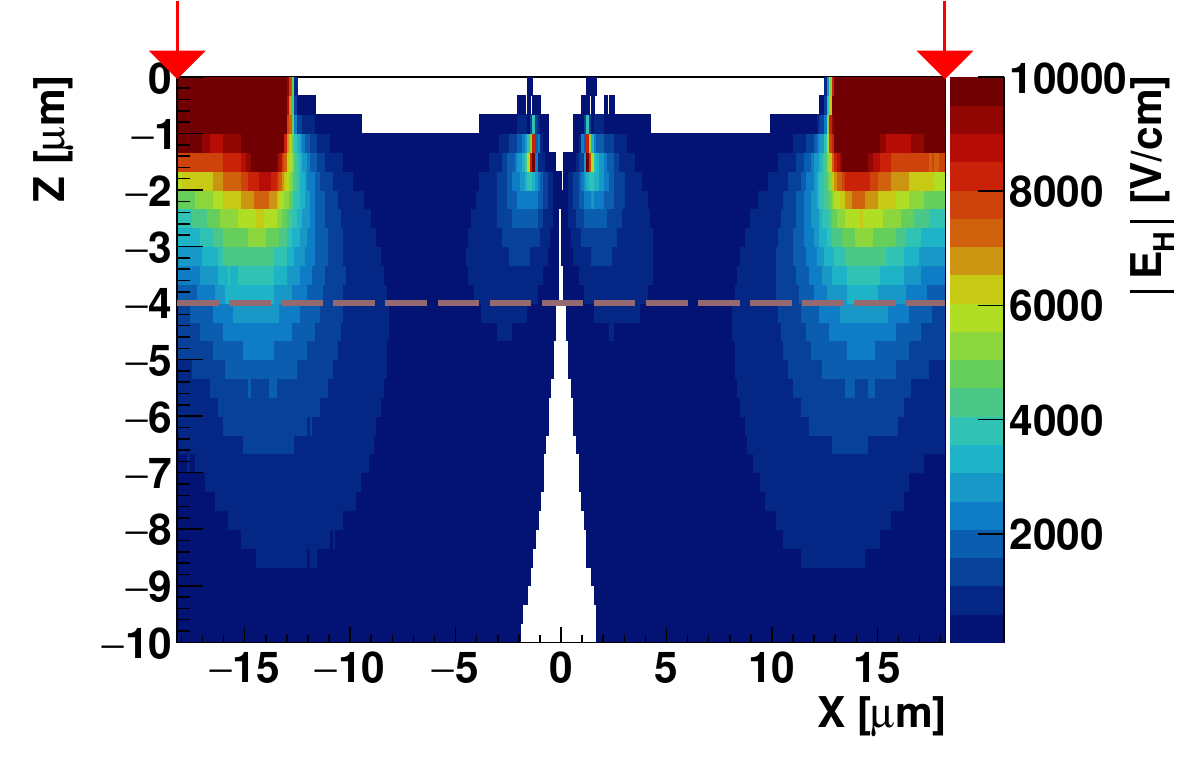}
    \label{fig:2defield:hor}
  }%
  \subfigure[]{
    \includegraphics[width=0.48\textwidth]{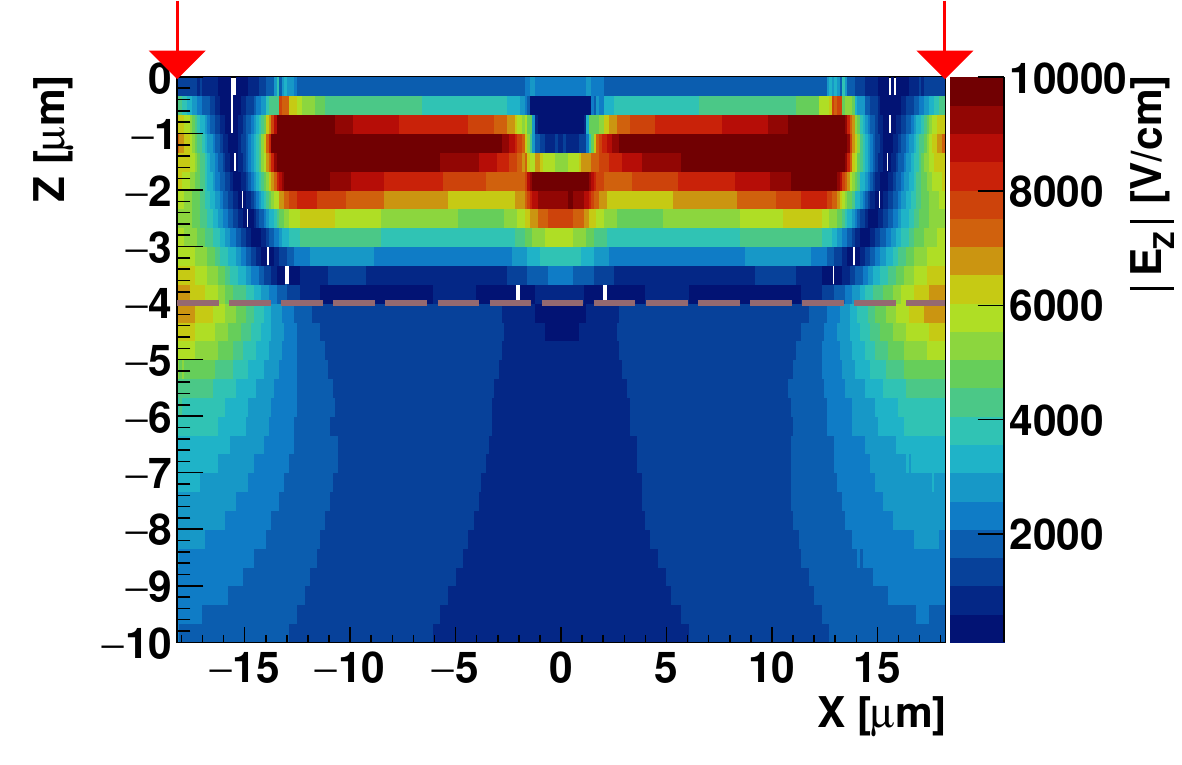}
    \label{fig:2defield:ver}
  }

  \subfigure[]{
    \includegraphics[width=0.48\textwidth]{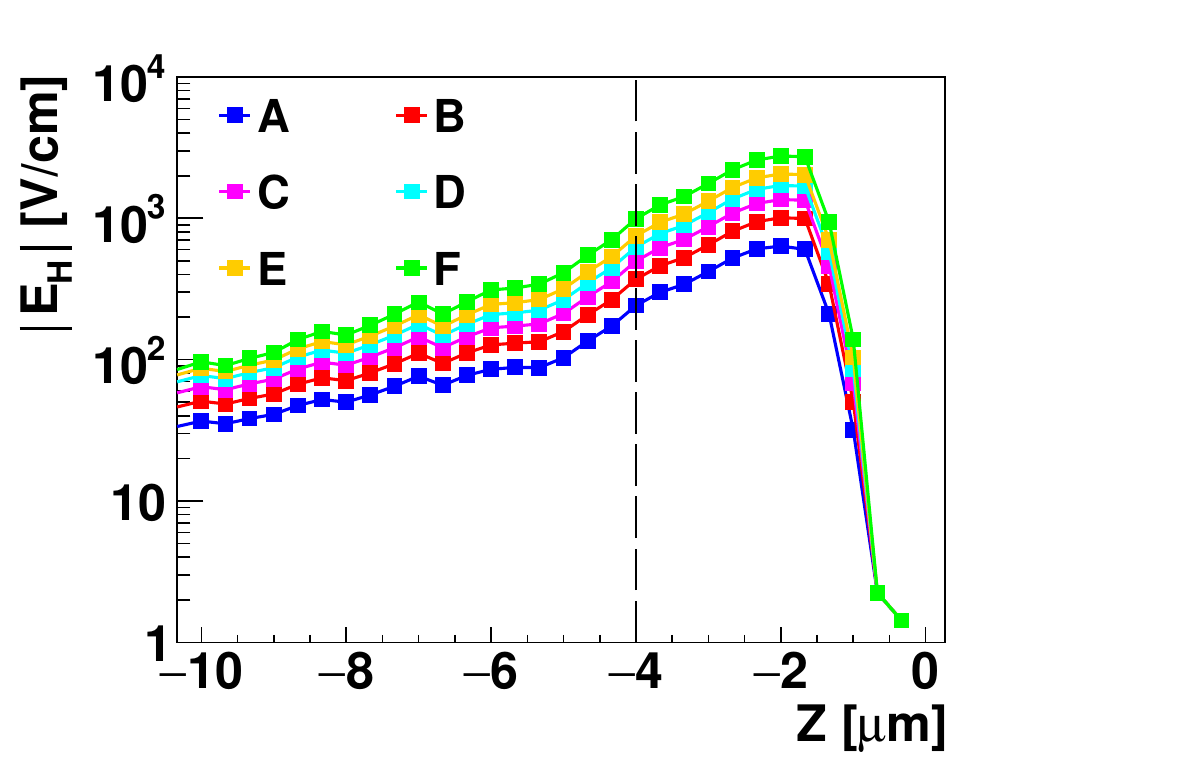}
    \label{fig:1defield:hor}
  }%
  \subfigure[]{
    \includegraphics[width=0.48\textwidth]{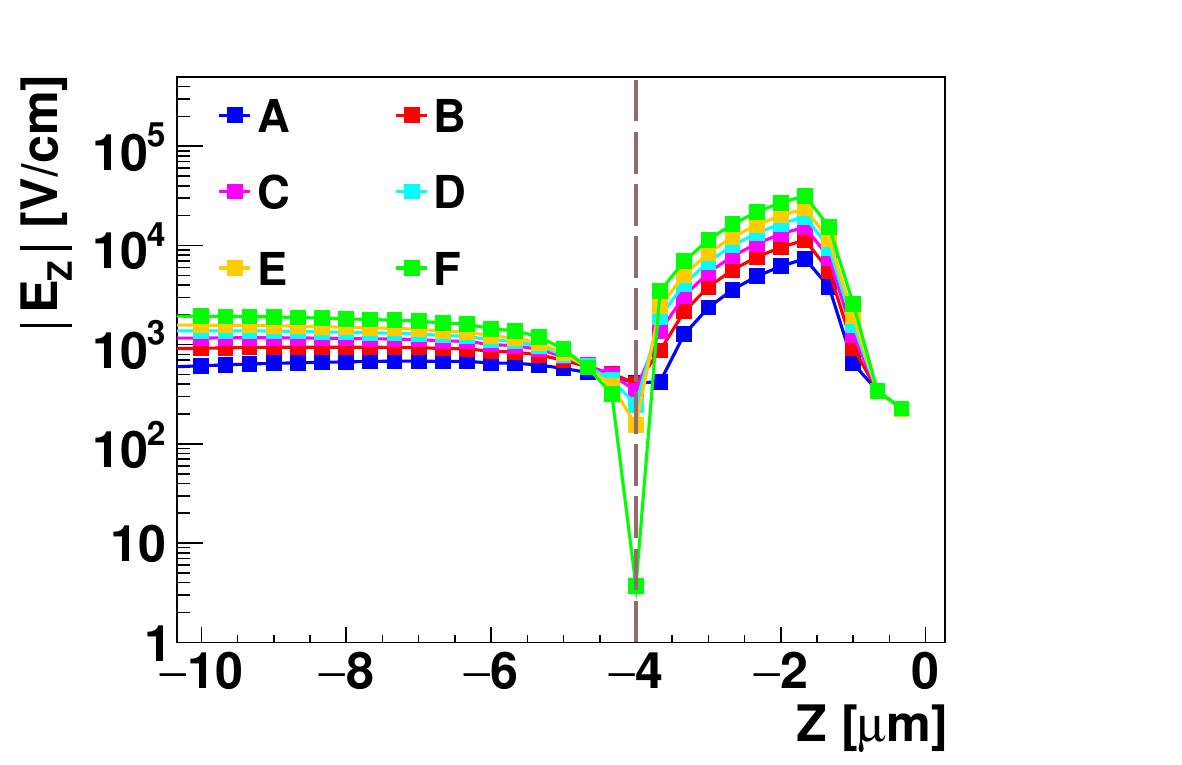}
    \label{fig:1defield:ver}
  }

 \caption{(a) Magnitudes of the horizontal and (b) vertical electric fields simulated in TCAD, shown across a cross-section from the 
 sensor surface to a depth of 10\,$\upmu$m at a bias voltage of -6\,V. The red arrow marks the position of the N-well diode. 
 (c) Horizontal and (d) vertical electric field magnitudes at the horizontal position $\text{X} \sim 0~\upmu\text{m}$, plotted as a function 
 of depth, for N-dop concentrations A-F. The brown dashed line in all panels indicates the interface between the N-blanket 
 and the P-type substrate.}
  \label{fig:2defield}
\end{figure}

These observations can be explained by the structure of the electric field depending on the N-dop concentration. 
\fig{2defield} (a) and (b) show the magnitudes of the horizontal and vertical electric fields, respectively, across a cross-section from 
the sensor surface to a depth of 10\,$\upmu$m, while (c) and (d) show the corresponding magnitudes at the pixel edge
(X $\sim 0~\upmu\text{m}$) as a function of depth. 
For hits at the pixel edge, the signal electrons generated in deep substrate (Z $\le -10~\upmu\text{m}$) initially propagate vertically 
toward the N-blanket due to the weak horizontal electric field ($\left|E_H \right| < 100~\text{V}/\text{cm}$, shown in (c)), and the 
comparatively stronger vertical electric field ($\left|E_Z \right| > 600~\text{V}/\text{cm}$, shown in (d)). Subsequently, 
they are separated by the stronger horizontal field at the N-blanket/substrate interface and drift toward the collection diodes. 
On one hand, at the pixel edge, as the N-dop concentration increases from A to F, a stronger horizontal electric field forms, as shown in 
(c), contributing to the improvement in charge collection and time response. On the other hand, at the same location, the magnitude 
of the vertical electric field (shown in (d)) increases with N-dop concentration except near the N-blanket/substrate interface, where it 
decreases gradually and falls below 10 V/cm for N-dop F. The increase in N-dop concentration enhances electrostatic screening, compressing 
the N-side depletion width and pinning the equipotential boundary near the P-well. The applied bias is thus predominantly dropped across 
the upper P-well/N-blanket junction, reducing the potential gradient at the lower boundary and leaving a vertically weak electric field 
region at the N-blanket/substrate interface. This low vertical electric field for N-dop beyond D hinders charge propagation, thereby 
degrading the charge collection and timing performance.

\subsection{Full telescope simulations}
\label{sect:allpix:fullsim}
The MALTA telescope \cite{MALTA_telescope}, along with the MALTA2 DUT sample, specified in \tab{telescope_geo}, was implemented in 
$\text{Allpix}^2$ for full-telescope simulations. For each plane, the sensor was modeled individually according to its specific
characteristics, including substrate type, sensor thickness, and process flavor.
The doping profiles and resulting electric field distributions were then imported into
$\text{Allpix}^2$. The responses of the telescope-plane models and the DUT models (with N-dop concentrations A to F) to the 120\,GeV/c 
proton beam were simulated, and the corresponding performance was analyzed after reconstruction in \texttt{MaltaTbAnalysis}. 

\begin{table}[htb!]
  \centering
  \caption{Specifications of the MALTA telescope planes and the MALTA2 DUT sample implemented in TCAD + $\text{Allpix}^2$.}
  \label{tab:telescope_geo}
  \begin{tabular}{rcccccccc}
    \hline 
    Plane & 1 & 2 & 3 & 4 & 5 & 6 & DUT \\
    \hline
    Sensor & \multicolumn{6}{c}{MALTA} & MALTA2 \\
    Matrix size & \multicolumn{6}{c}{512$\times$512} & 224$\times$512 \\
    \hline
    Pixel pitch ($\upmu\text{m}^{2}$) & \multicolumn{7}{c}{$36.4\,\times\,36.4$} \\ 
    \hline
    Substrate & EPI & Cz & Cz & Cz & Cz & EPI & Cz \\
    Flavor & STD & NGAP & STD & STD & NGAP & STD & XDPW \\
    Thickness ($\upmu$m) & 100 & 100 & 300 & 300 & 100 & 300 & 100\\
    Operating voltage (V) & -6 & -6 & -30 & -30 & -6 & -6 & -6 \\
    Position in beam direction (mm) & 0 & 80 & 160 & 940 & 1020 & 1100 & 642 \\
    \hline
  \end{tabular}
\end{table}

\begin{figure}[htb!]
  \centering
  \subfigure[]{
    \includegraphics[width=0.32\textwidth]{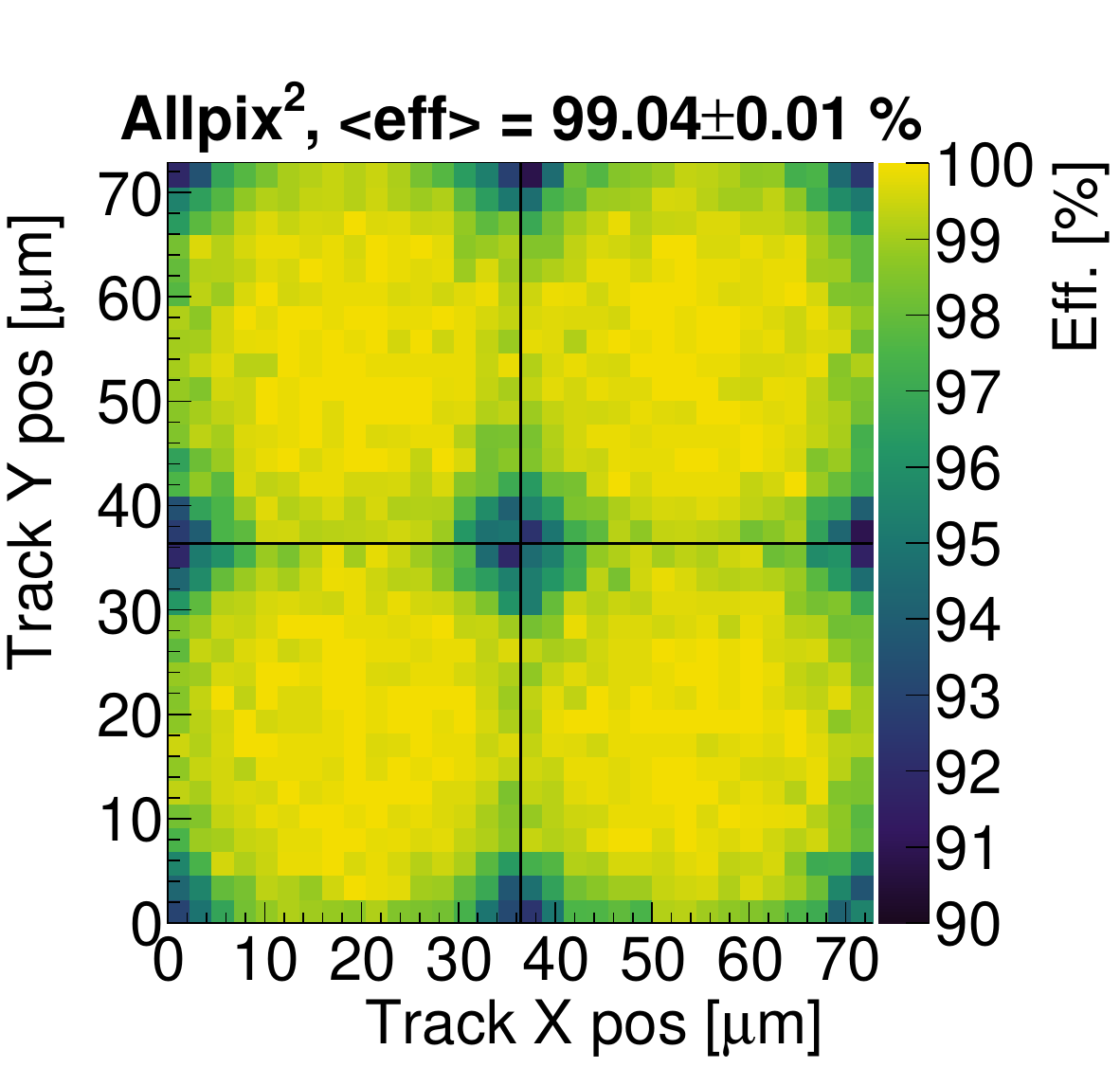}
    \label{fig:inpixsim:eff2e14}
  }%
  \subfigure[]{
    \includegraphics[width=0.32\textwidth]{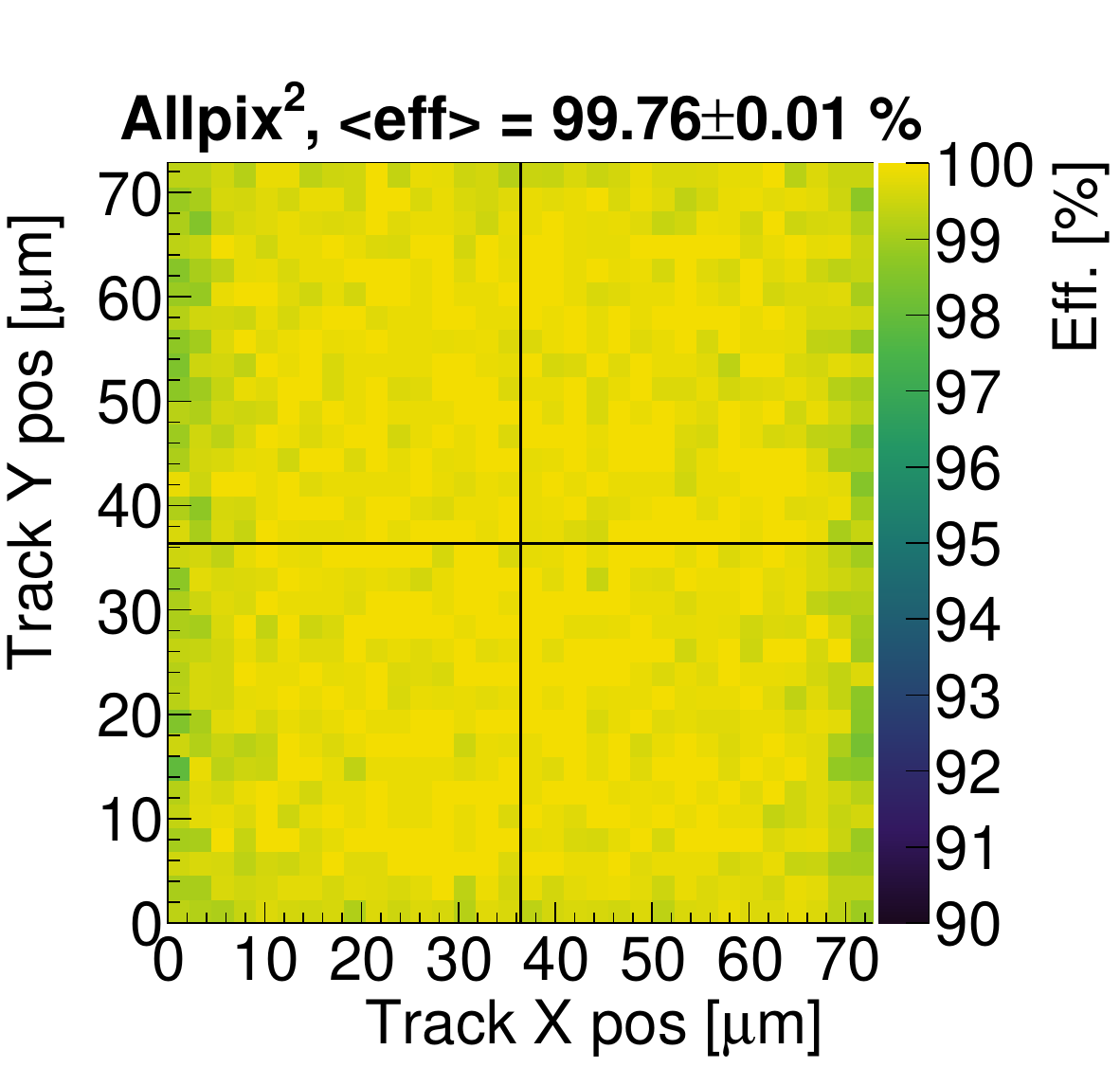}
    \label{fig:inpixsim:eff5e14}
  }%
  \subfigure[]{
    \includegraphics[width=0.32\textwidth]{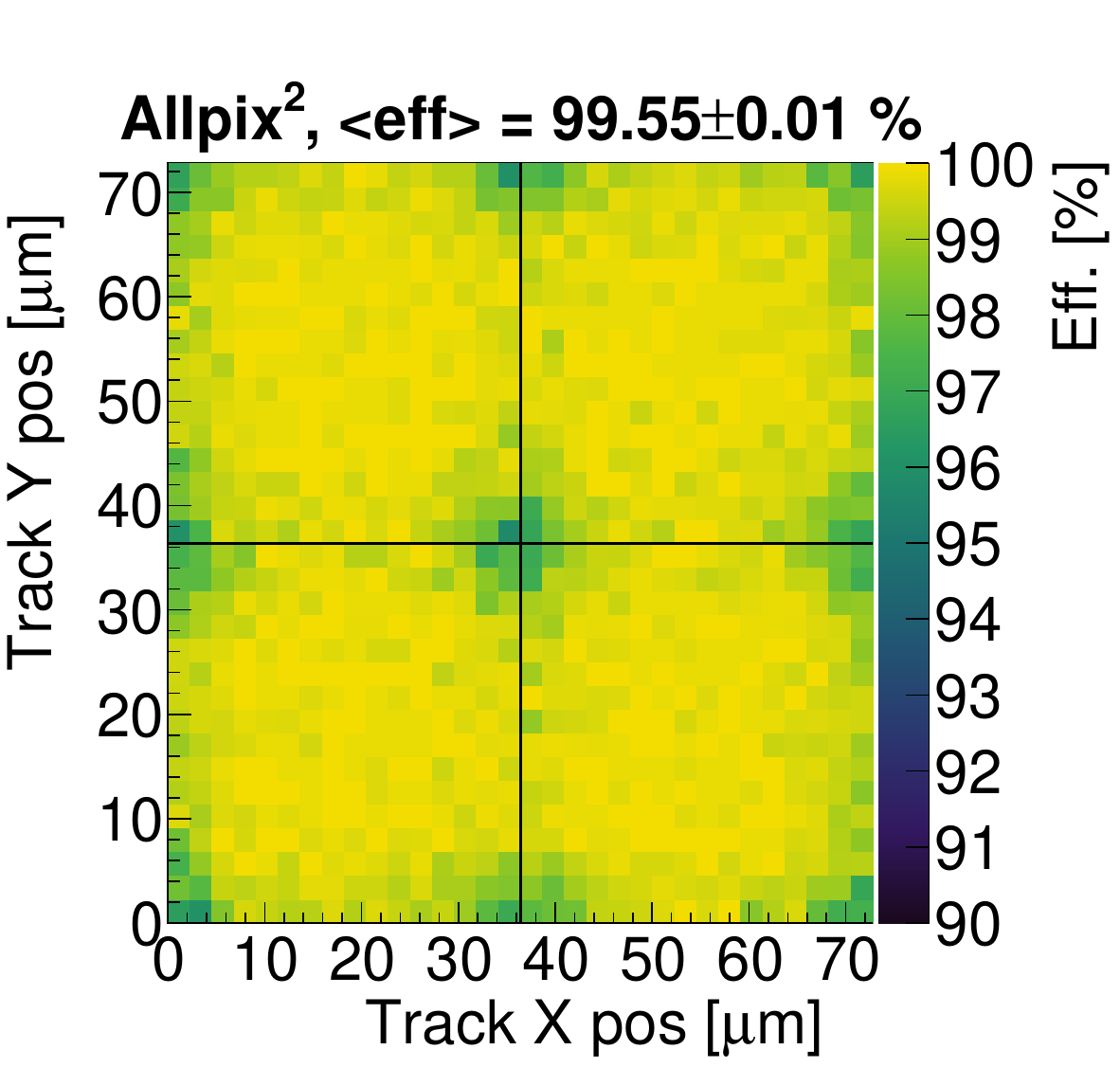}
    \label{fig:inpixsim:eff8e14}
  }

  \subfigure[]{
    \includegraphics[width=0.32\textwidth]{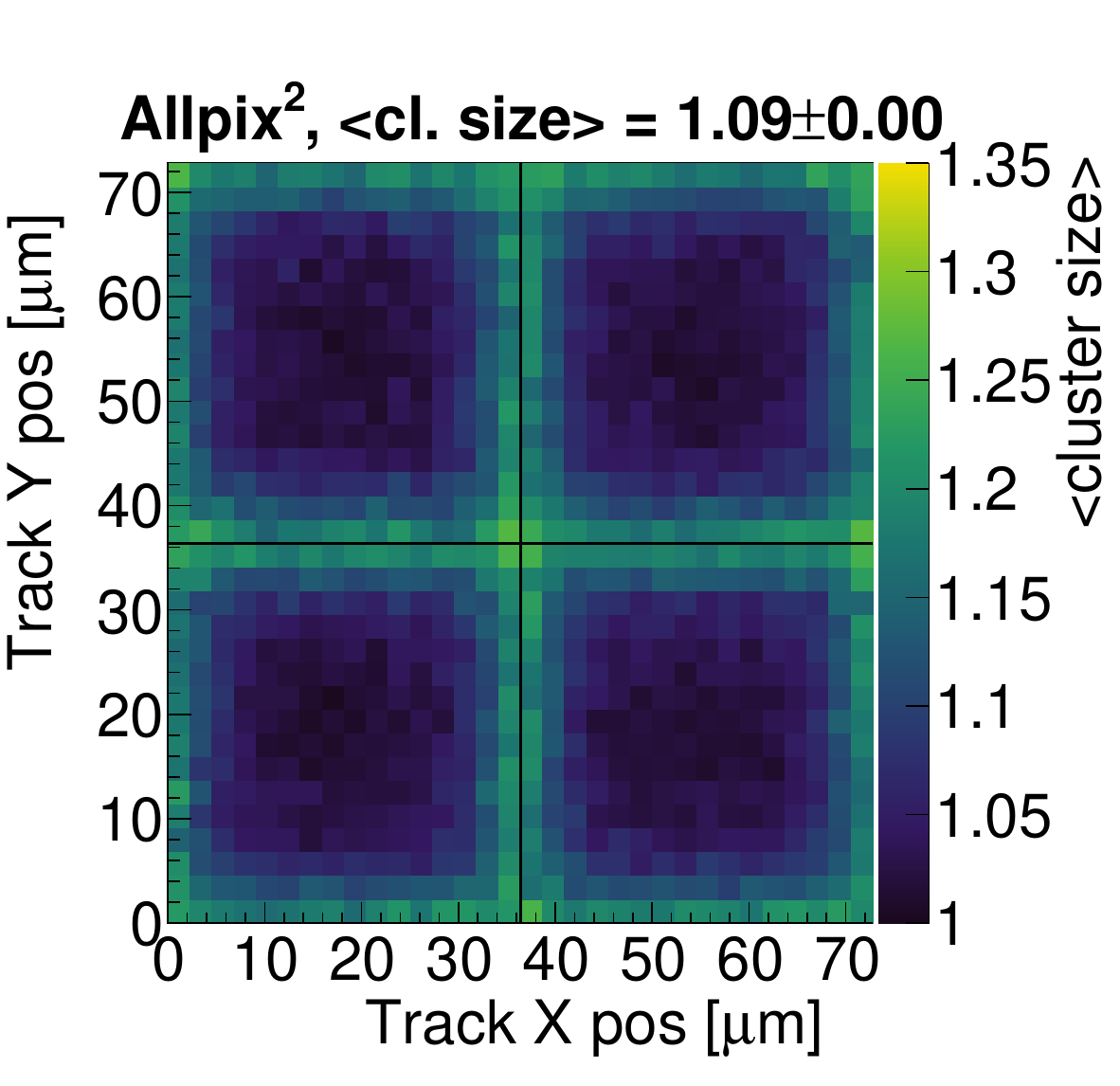}
    \label{fig:inpixsim:clsize2e14}
  }%
  \subfigure[]{
    \includegraphics[width=0.32\textwidth]{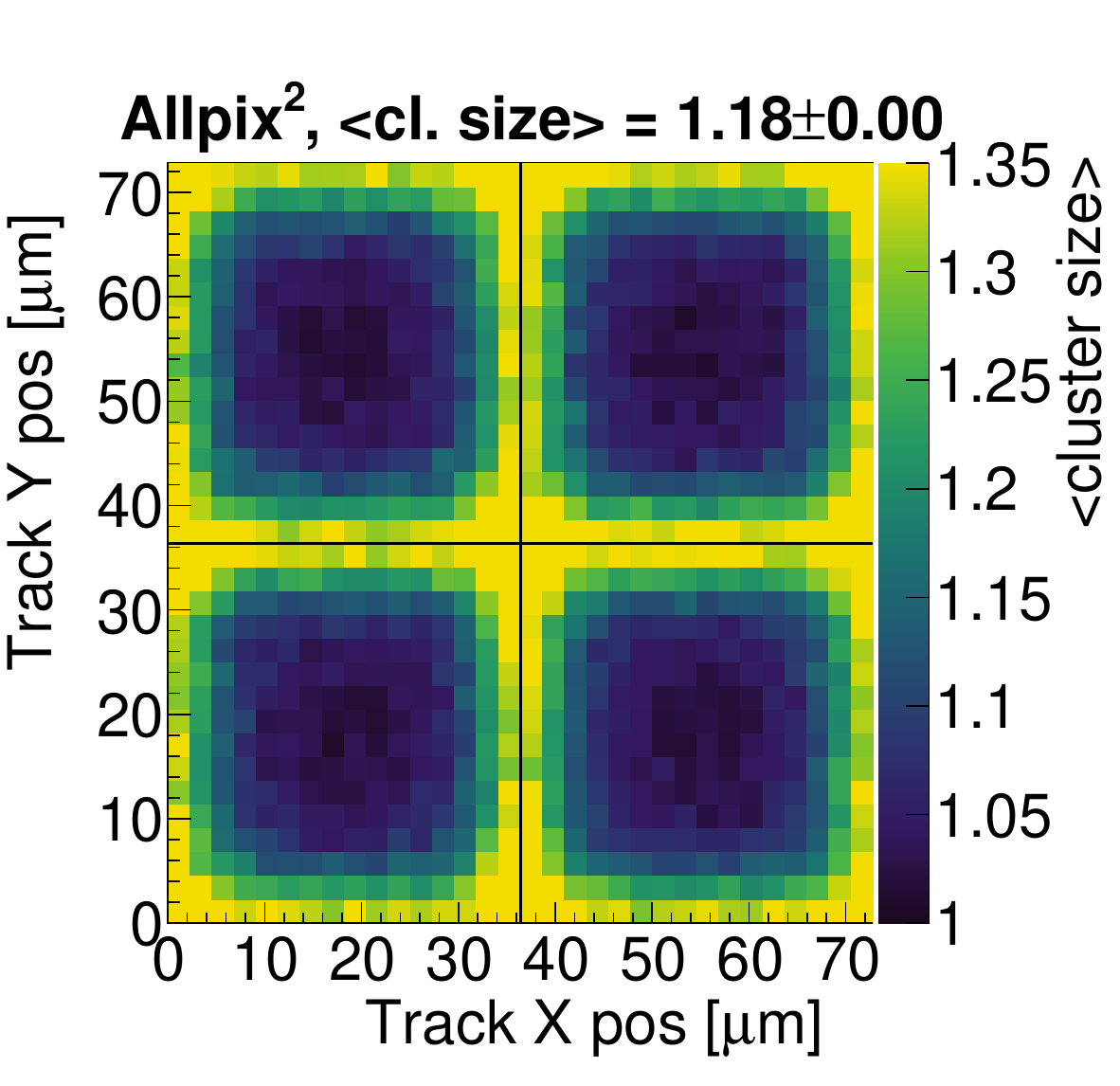}
    \label{fig:inpixsim:clsize5e14}
  }%
  \subfigure[]{
    \includegraphics[width=0.32\textwidth]{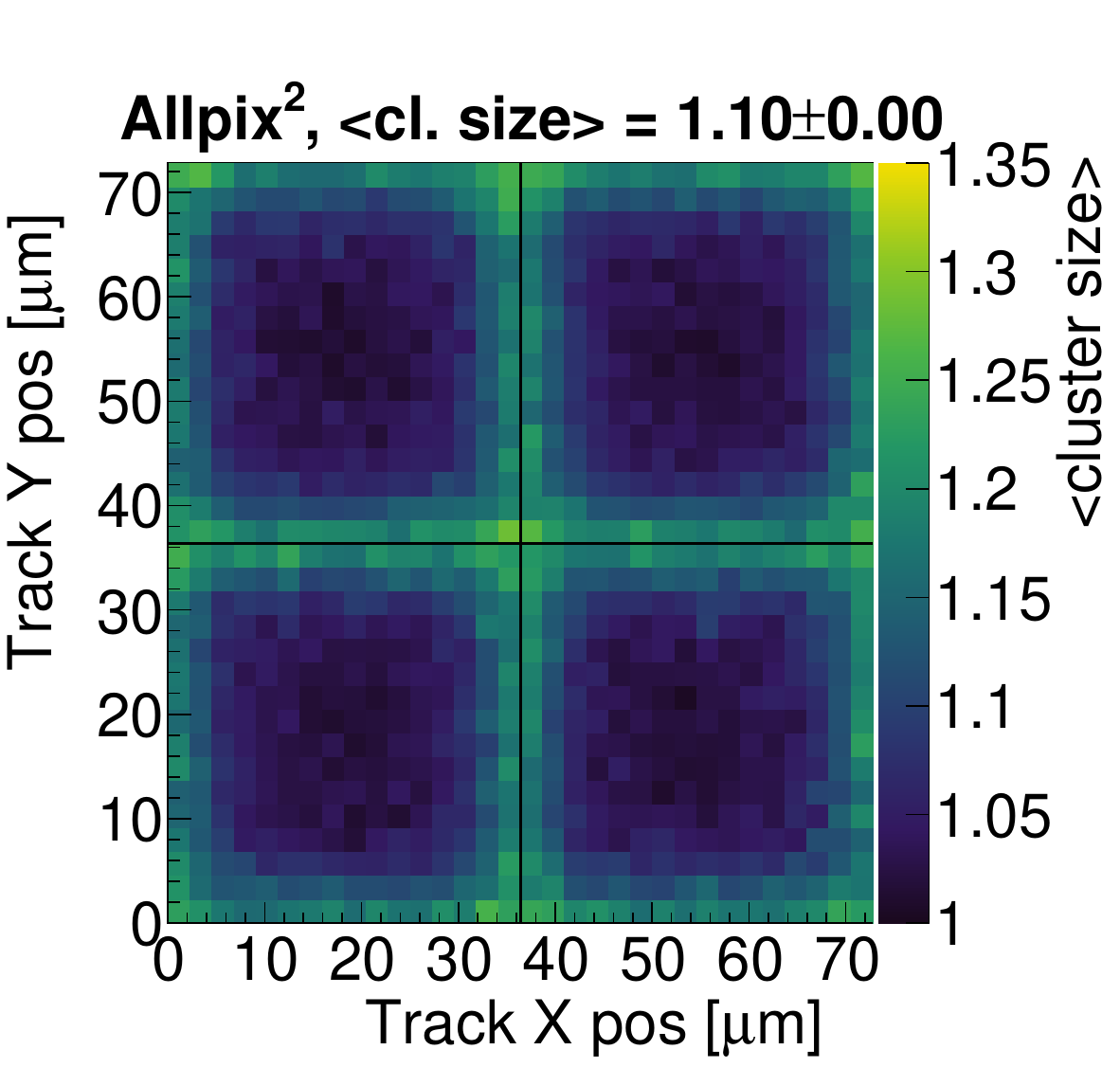}
    \label{fig:inpixsim:clsize8e14}
  }

  \subfigure[]{
    \includegraphics[width=0.32\textwidth]{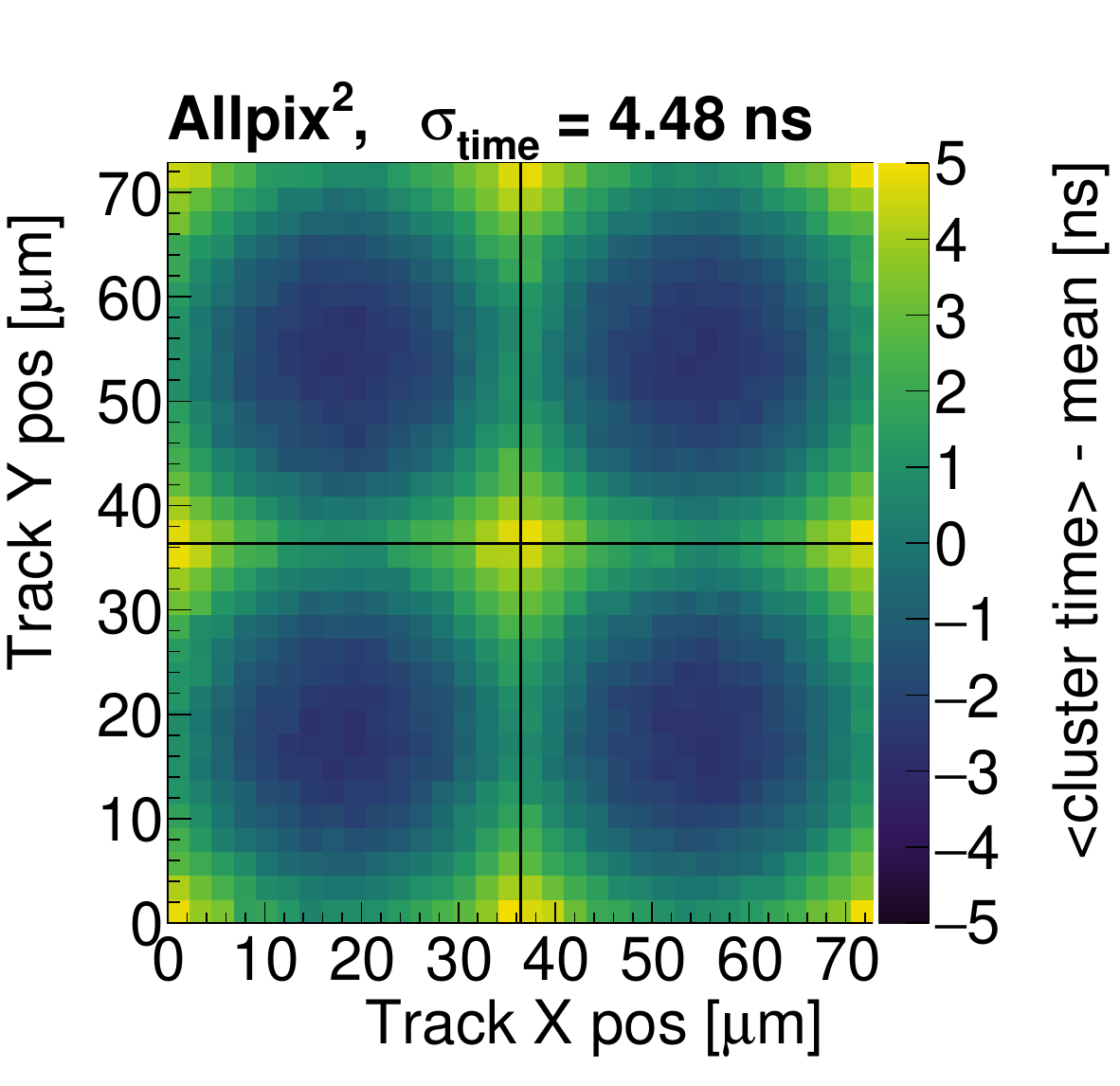}
    \label{fig:inpixsim:cltime2e14}
  }%
  \subfigure[]{
    \includegraphics[width=0.32\textwidth]{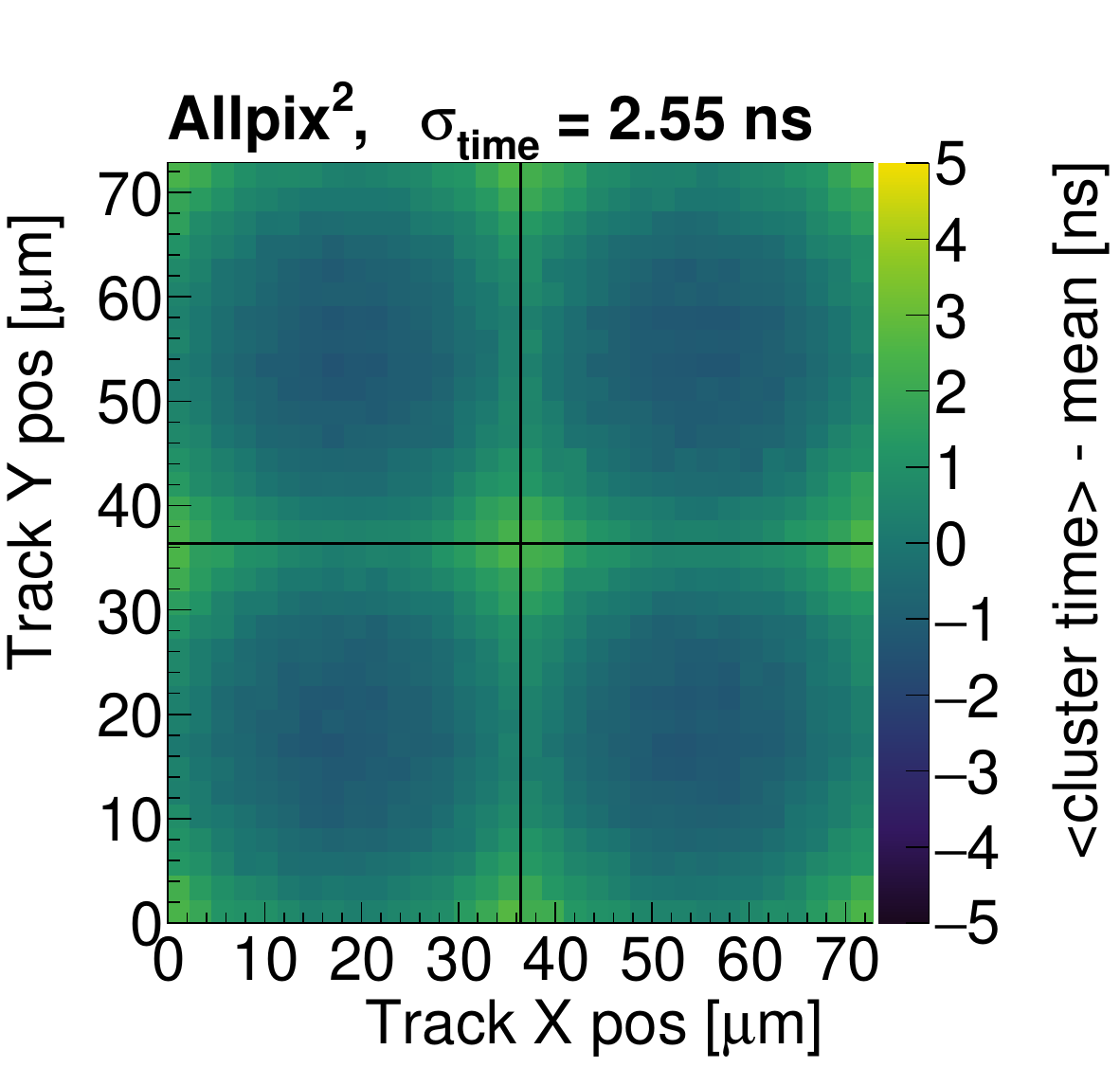}
    \label{fig:inpixsim:cltime5e14}
  }%
  \subfigure[]{
    \includegraphics[width=0.32\textwidth]{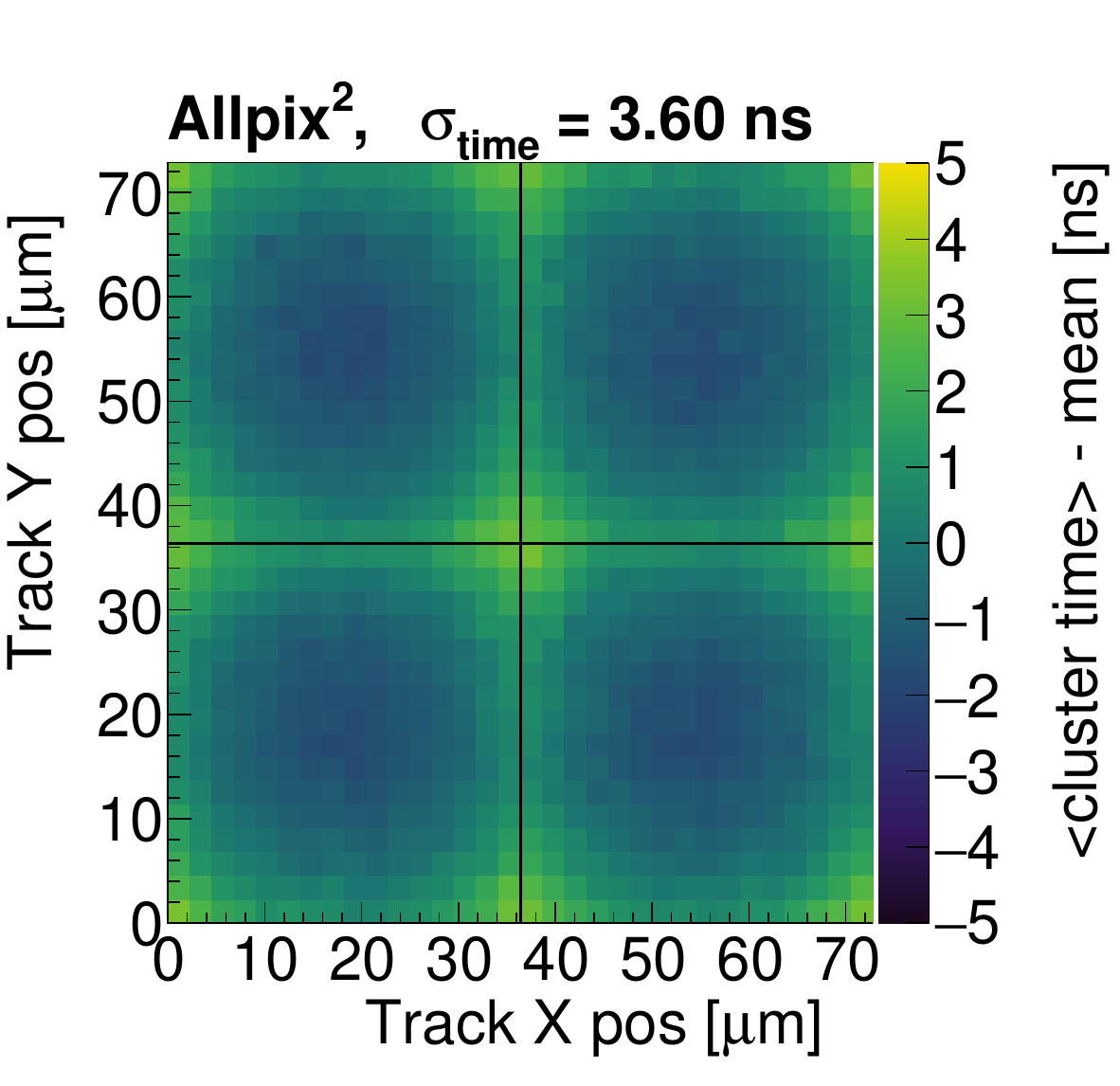}
    \label{fig:inpixsim:cltime8e14}
  }
  \caption{The in-pixel 2D detection efficiency (a-c), cluster size (d-f) and the difference between the timing of the leading hit in the 
  cluster and the average timing over the entire matrix (g-i) with N-dop concentrations, from left to right: A, D 
  and F. The bias voltage is set to -6\,V and the threshold is 650\,$\text{e}^-$.}
  \label{fig:inpixsim}
\end{figure}

The in-pixel detection efficiency, cluster size and timing performance of MALTA2 for the N-dop concentrations A, D and F are shown in \fig{inpixsim}. 
For the lowest concentration (A), a structured detection efficiency with an average value of approximately 99\% is observed 
(\fig{inpixsim:eff2e14}), reaching nearly 100\% over most of the pixel area and decreasing only at the edges. \fig{inpixsim:clsize2e14} 
presents the corresponding 2D cluster-size distribution for the same N-dop value. Single-pixel clusters dominate across the pixel, with 
charge sharing occurring mainly at the edges, resulting in an average cluster size of about 1.09 pixels. The timing difference between 
the leading hit and the average timing across the matrix, shown in \fig{inpixsim:cltime2e14}, exhibits a distribution similar to 
that of the cluster size, with a spread of approximately 4.48 ns across the pixel. As N-dop increases from A to D, higher and more 
uniform detection efficiencies are observed. Simultaneously, the cluster size increases 
and the timing spread decreases, indicating improved charge propagation. However, further increasing N-dop beyond D leads to degraded 
sensor performance, consistent with the observations in \fig{dut} and attributable to the electric field dependence on N-dop concentration 
at the pixel edges shown in \fig{2defield}.

\subsubsection{Comparisons with test-beam measurements}
\label{sect:allpix:valid}
The simulated sensor performance was compared with test-beam measurements obtained with a 120\,GeV/c hadron beam at the CERN SPS in 2025. 
The detection efficiency, cluster size, and ToA of the leading hit were compared to validate the simulations and to determine the optimal N-dop
 concentration for the configurations used in the grazing-angle simulation (\sect{allpix:ga}) which studies the active depth of the depleted
region of the MALTA2 sensor.

\begin{figure}[htb!]
  \centering
  \subfigure[]{
    \includegraphics[width=0.48\textwidth]{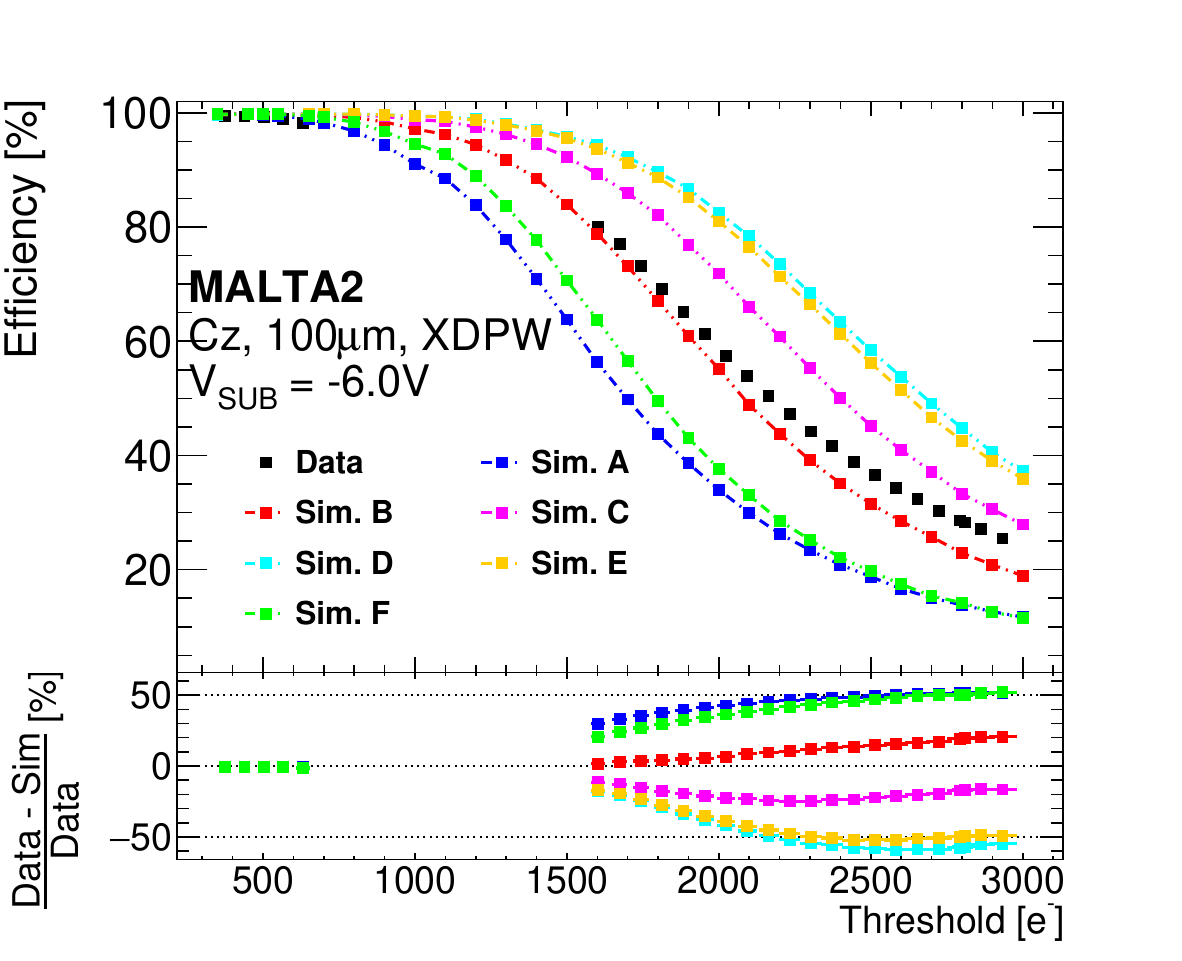}
    \label{fig:thrscan:eff}
  }
  \subfigure[]{
    \includegraphics[width=0.48\textwidth]{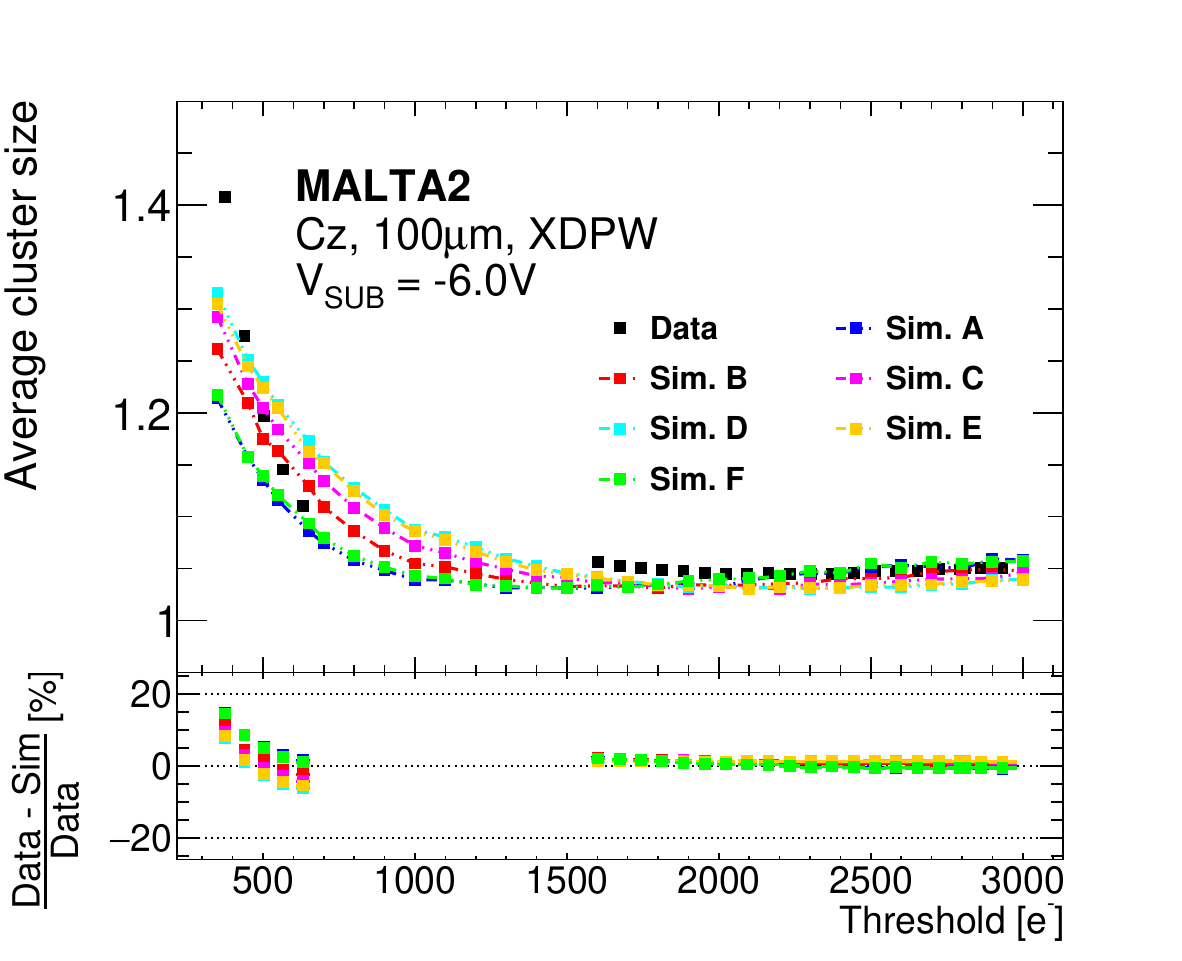}
    \label{fig:thrscan:clsize}
  }
  \caption{(a) Detection efficiencies and (b) cluster sizes for the measurements and simulations with N-dop concentrations A to F, 
  as a function of threshold. The relative residuals are derived via linear interpolation between the two nearest simulated points adjacent 
  to each measurement.}
  \label{fig:thrscan}
\end{figure}
The detection efficiency and the cluster size comparisons between measurements and simulations with N-dop concentrations A to F as a function of
the threshold are shown in \fig{thrscan} (a) and (b), respectively. In measurements, the sensor under test was 
configured in different gain modes \cite{MALTA2_ChargeCalibration} to achieve a wide range of thresholds from 350 to 3000\,$\text{e}^-$. 
A gap in threshold coverage, ranging from 700 to 1600\,$\text{e}^-$, is observed in measurements between two gain modes. 

For the detection efficiency, agreement is found in the low threshold region ($< 700\,\text{e}^-$), where the relative spread among the
 simulated efficiencies for different N-dop concentrations is within 2\% of the measured values. At higher thresholds ($> 1600\,\text{e}^-$),
 however, the simulated efficiencies diverge significantly, with deviations reaching up to 52\%. Among all N-dop concentrations, B and C 
 show the closest agreement with the measured values, though in opposite directions: B underestimates the efficiency by up to 22\%, while 
 C overestimates it by up to 25\%.

The cluster size exhibits a different threshold dependence. Close agreement is observed at high thresholds ($> 1600\,\text{e}^-$), where
single-pixel clusters dominate and the relative variations are less than 3\%. In the low-threshold region ($< 700\,\text{e}^-$), the
simulations show larger discrepancies and underestimate the cluster size by up to 15\% at 350\,$\text{e}^{-}$. Moreover, 
the measured cluster size drops more steeply 
as the threshold increases from 350 to 650\,$\text{e}^{-}$, suggesting that further optimizations of the input parameters, particularly 
all the doping profiles, are needed. Again, N-dop concentrations B and C yield better agreement, with smaller relative deviations from the 
measured cluster size across all thresholds.

\begin{figure}[htb!]
  \centering
  \subfigure[]{
    \includegraphics[width=0.45\textwidth]{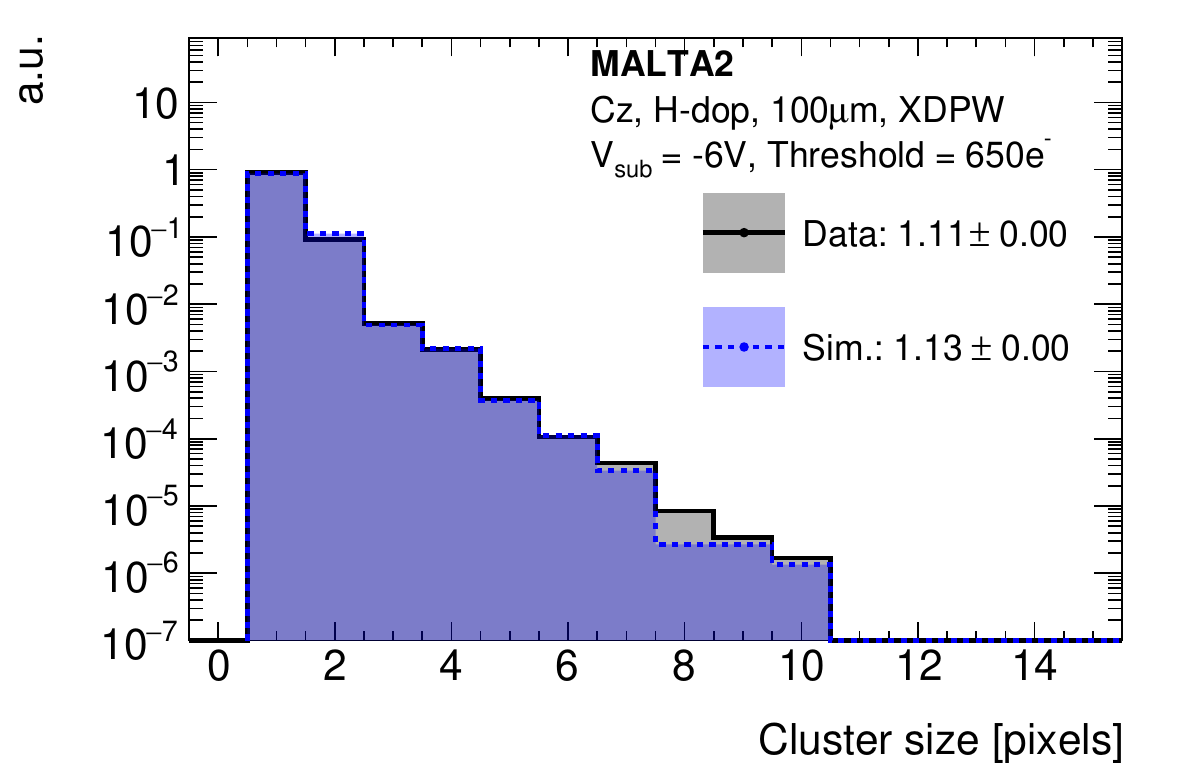}
    \label{fig:cmp:clsize650}
  }
  \subfigure[]{
    \includegraphics[width=0.45\textwidth]{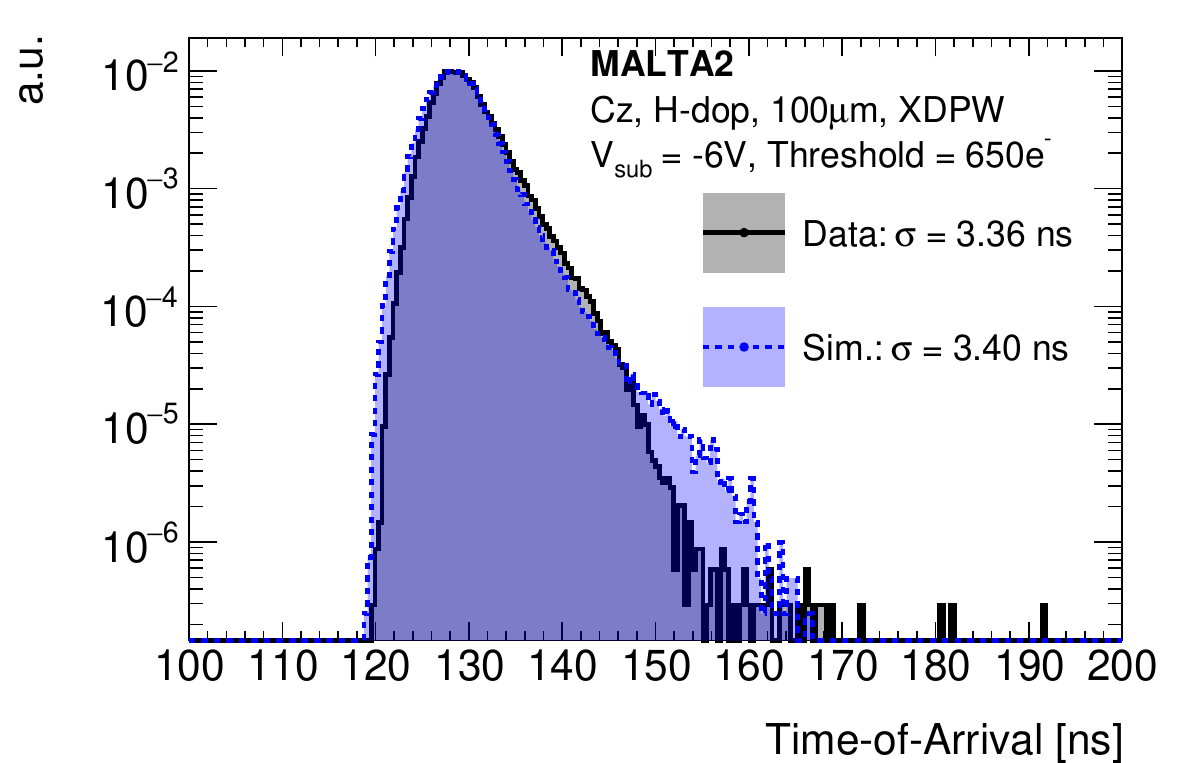}
    \label{fig:cmp:toa650}
  }

  \caption{Distributions of (a) cluster size and (b) leading-hit ToA for the measurement (black) and simulation (blue). 
  The simulated ToA is peak-aligned to the measurement to account for the external delay in the measurement. The bias voltage is set to -6\,V
  and the threshold is 650\,$\text{e}^-$.}
  \label{fig:cmp}
\end{figure}

\fig{cmp} (a) and (b) compare the distributions of cluster size and leading-hit ToA between
 measurement and simulation at a threshold of 650\,$\text{e}^-$, respectively. The average cluster size in measurement is 
 approximately 1.11, while the simulation records a value of approximately 1.13, showing a quantitative match with a relative 
 residual below 2\%. In addition, the measured and simulated standard deviations of the ToA distributions are approximately 
 3.36\,ns and 3.40\,ns, agreeing within 2\%.

\begin{figure}[htb!]
  \centering
     \subfigure[]{
    \includegraphics[width=0.32\textwidth]{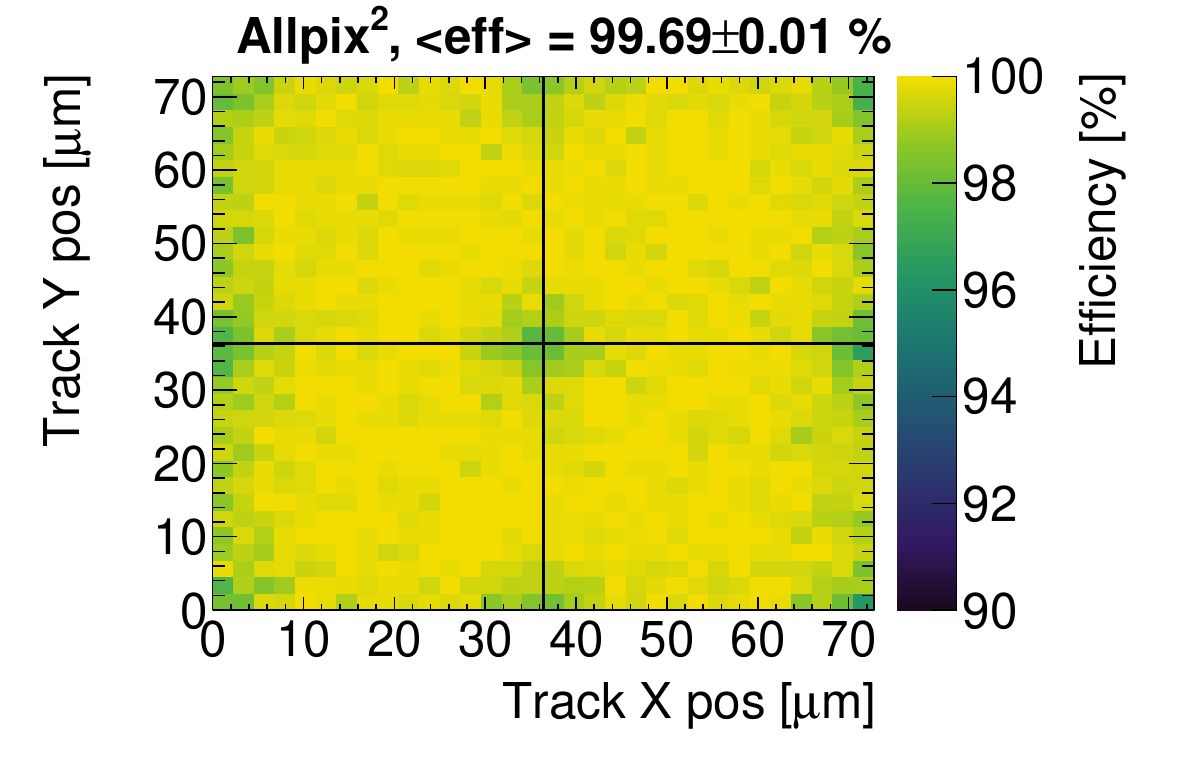}
    \label{fig:inpixres:effsim}
  }%
   \subfigure[]{
    \includegraphics[width=0.32\textwidth]{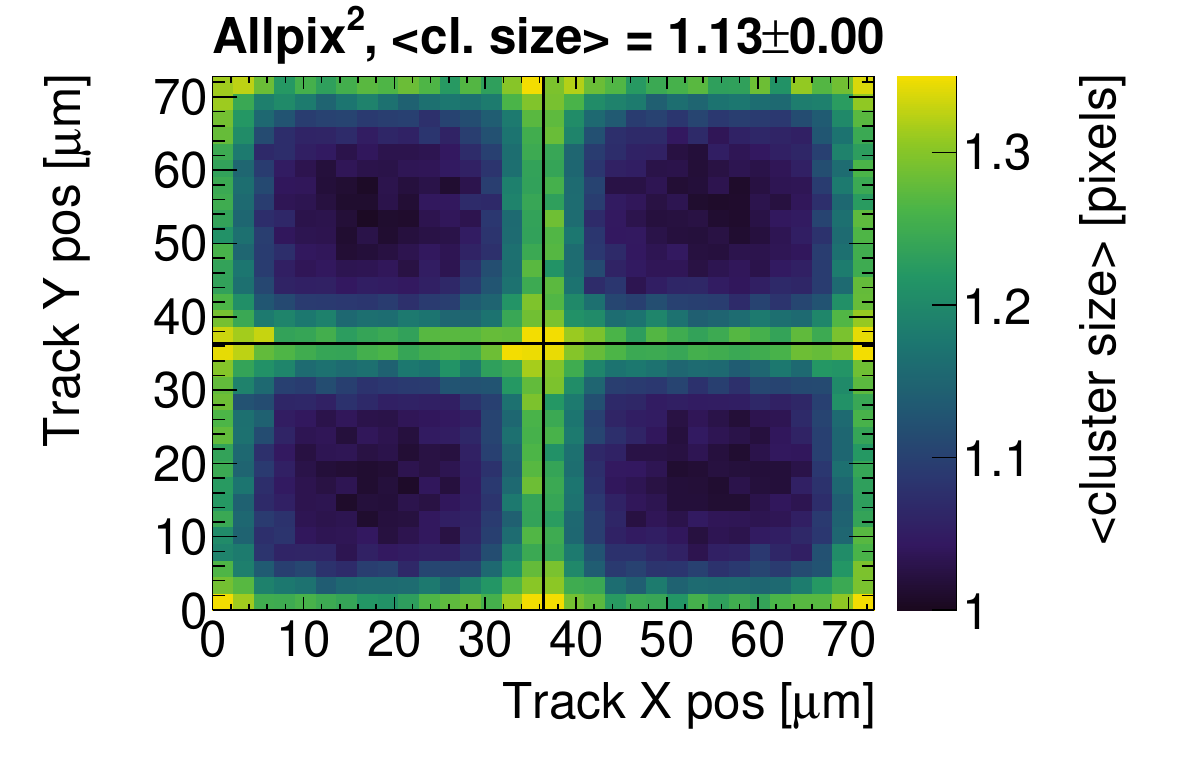}
    \label{fig:inpixres:clsizesim}
  }%
  \subfigure[]{
    \includegraphics[width=0.32\textwidth]{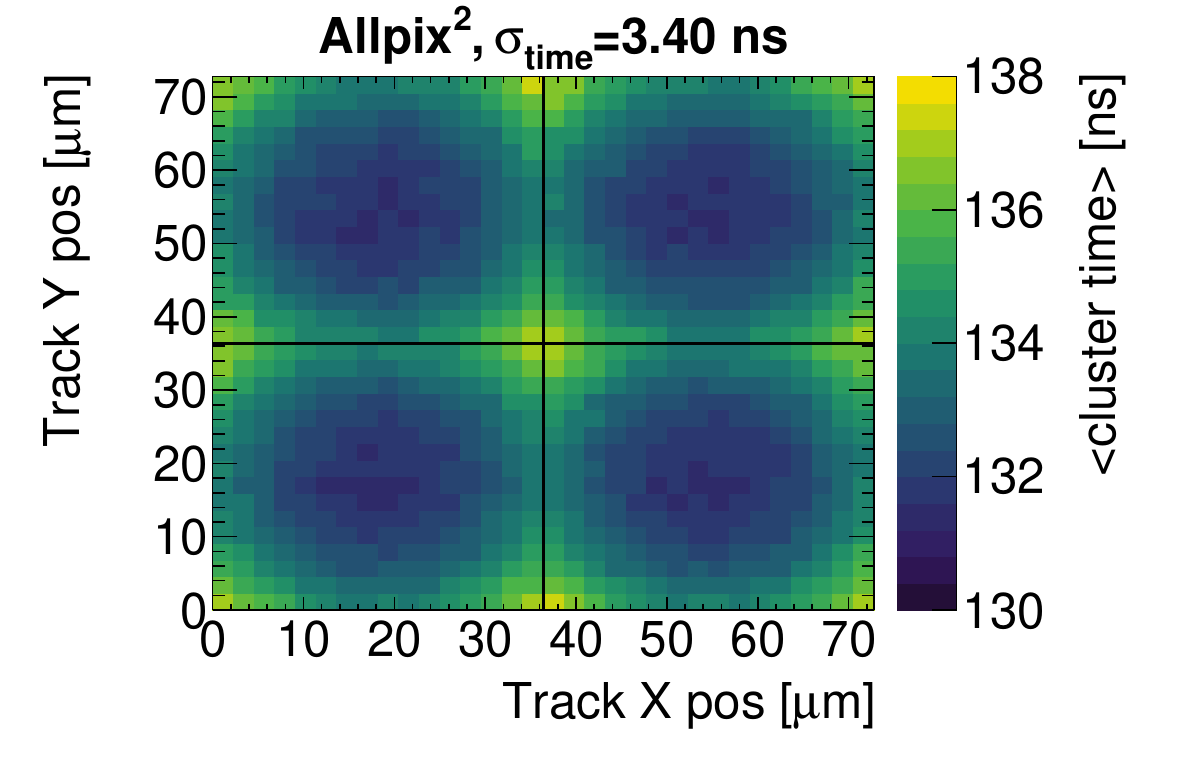}
    \label{fig:inpixres:cltimesim}
  }
    \subfigure[]{
    \includegraphics[width=0.32\textwidth]{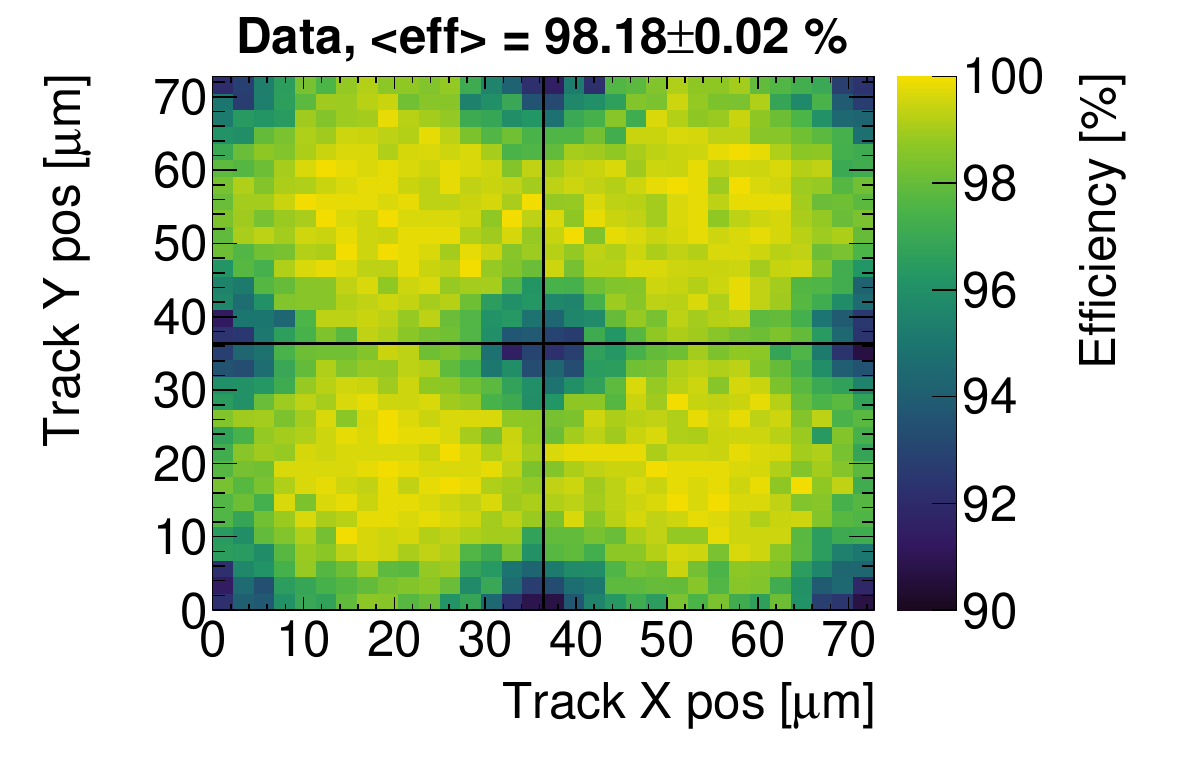}
    \label{fig:inpixres:effmea}
  }%
    \subfigure[]{
    \includegraphics[width=0.32\textwidth]{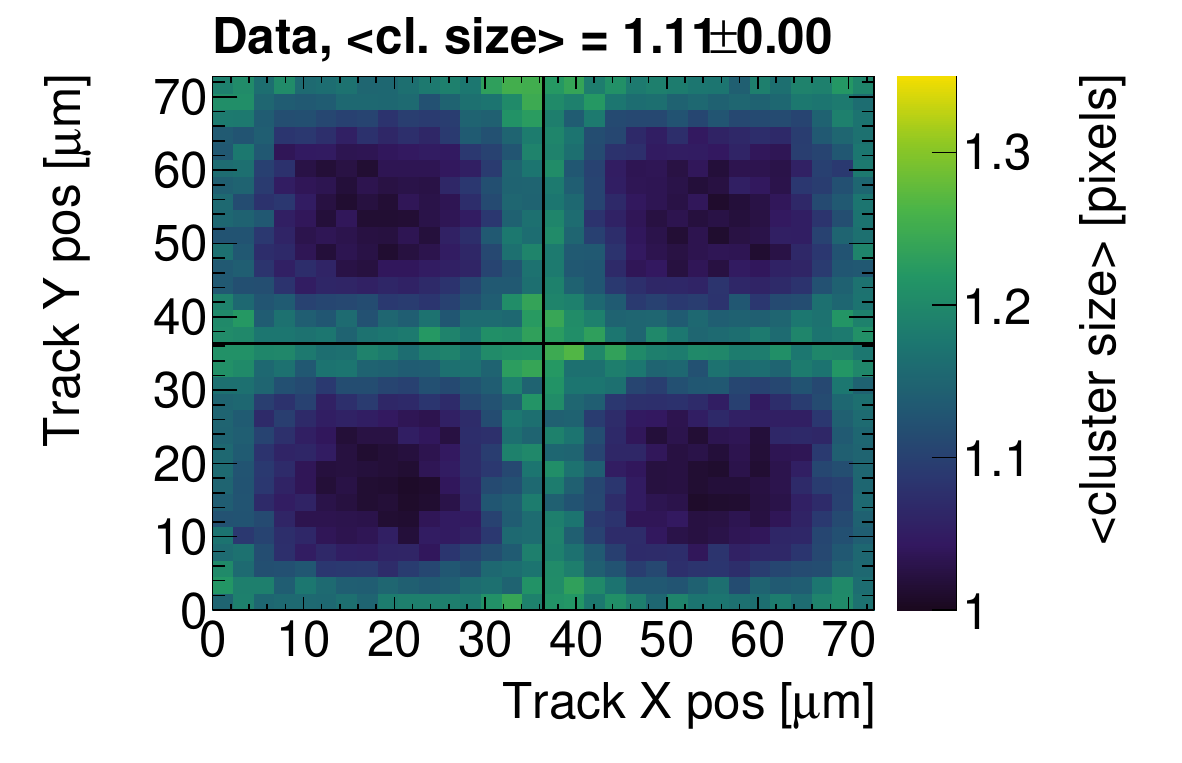}
    \label{fig:inpixres:clsizemea}
  }%
  \subfigure[]{
    \includegraphics[width=0.32\textwidth]{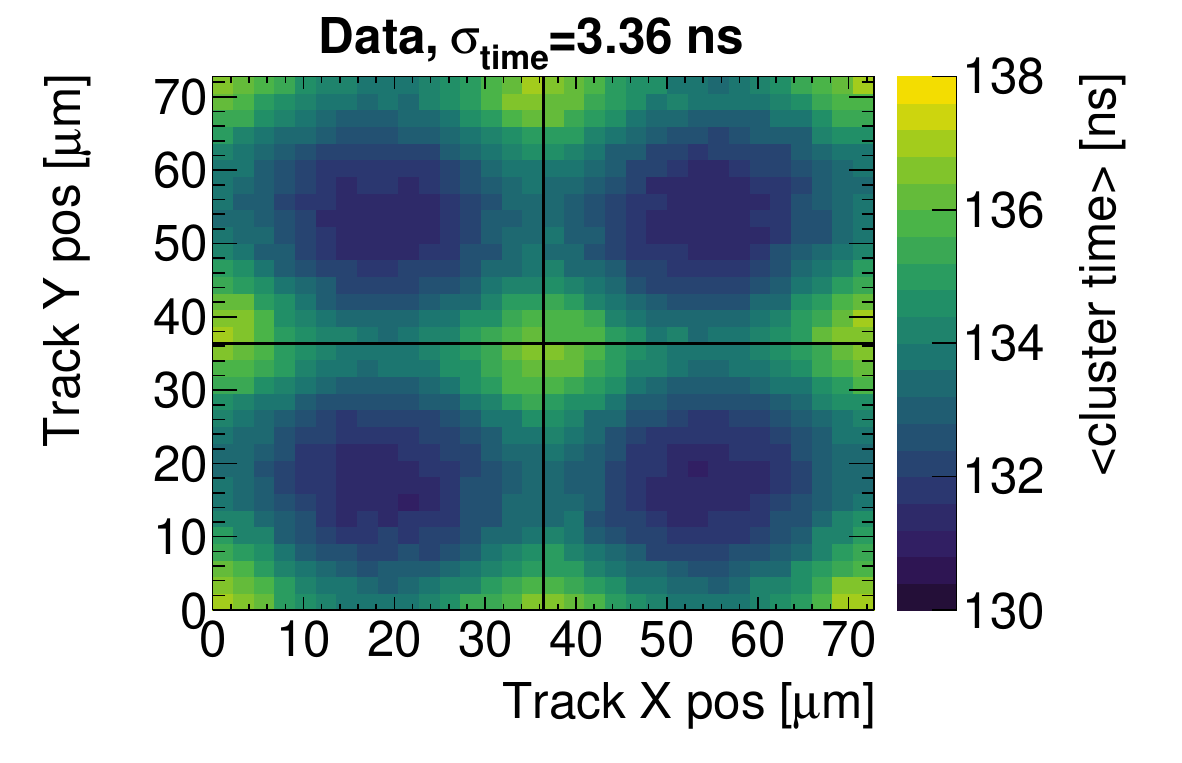}
    \label{fig:inpixres:cltimemea}
  }
  \subfigure[]{
    \includegraphics[width=0.32\textwidth]{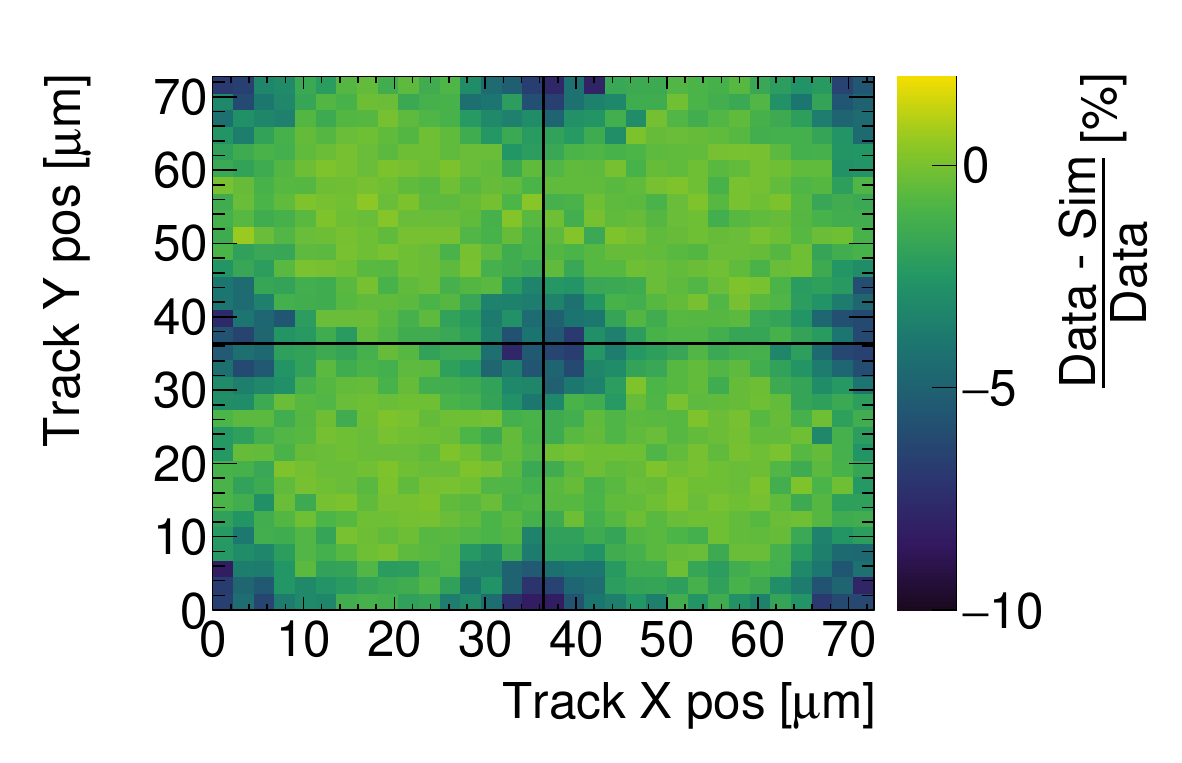}
    \label{fig:inpixres:effcmp}
  }%
   \subfigure[]{
    \includegraphics[width=0.32\textwidth]{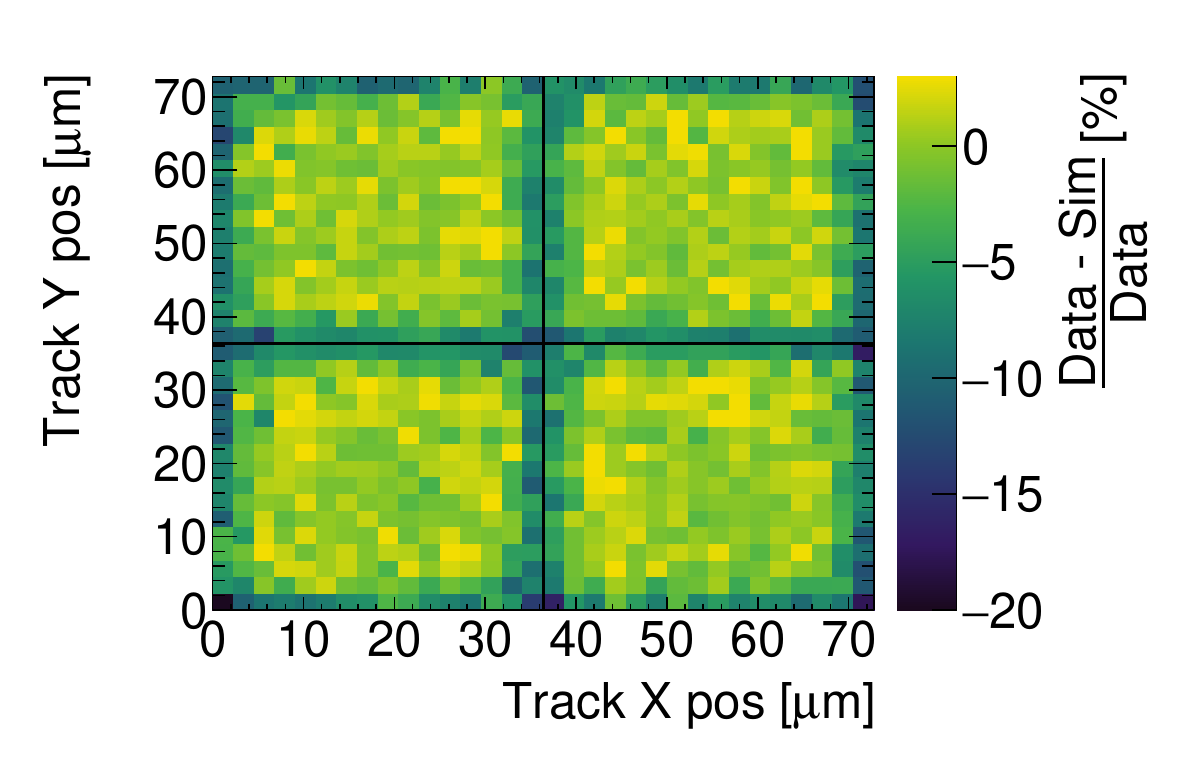}
    \label{fig:inpixres:clsizecmp}
  }%
  \subfigure[]{
    \includegraphics[width=0.32\textwidth]{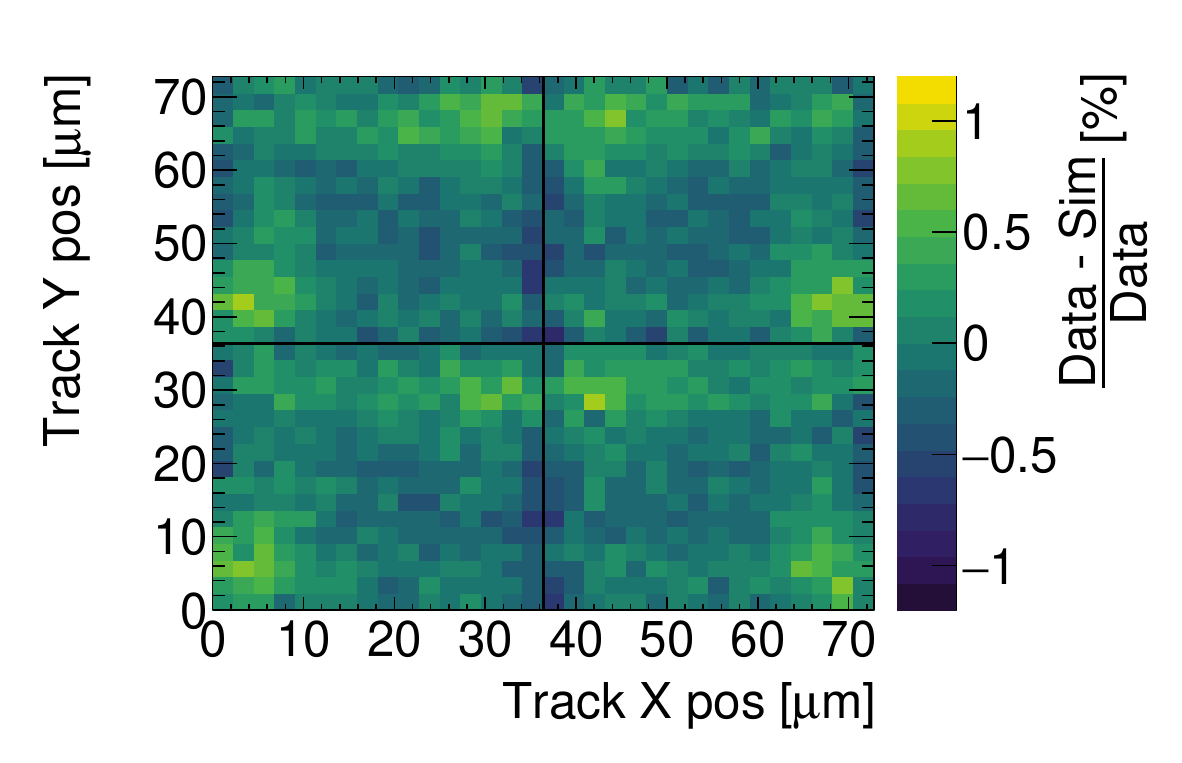}
    \label{fig:inpixres:cltimecmp}
  }
  \caption{
  (a) Simulated and (d) measured in-pixel detection efficiency, and (g) their relative residual.
  (b) Simulated and (e) measured in-pixel cluster size, and (h) their relative residual.
  (c) Simulated and (f) measured in-pixel ToA and (i) their relative residual.  The same 
  peak-alignment procedure applied to the 1D distributions in \fig{cmp:toa650} 
  (to correct for the external delay in measurements) is used here.
  The bias voltage is -6\,V and the threshold is 650\,$\text{e}^{-}$.
  }
  \label{fig:inpixres}
\end{figure}

The in-pixel distributions of detection efficiency, cluster size, and leading-hit ToA from simulation and measurement are shown 
in \fig{inpixres} (a)--(f), with their relative residuals shown in (g)--(i), respectively. 
Overall, the simulation reproduces the main spatial features observed in the measurement: high efficiency, small cluster size, 
and short ToA are found in the pixel center, while the edges show degraded efficiency and time response with larger cluster 
size due to charge sharing.
The residual maps show, however, that the agreement is not uniform across the 
pixel. The relative residual of efficiency is below 3\% over most of the pixel area, with an overestimation of up to 8\% at the pixel edges. 
The cluster size residuals behave similarly, with a maximum discrepancy of 15\% at the edges. The ToA residuals are within 1\% across 
the entire pixel. The observed in-pixel discrepancies, especially at the pixel edges, suggest a lack of realistic description of 
the electric field.

\subsubsection{Grazing-angle studies \& active depth estimation}
\label{sect:allpix:ga}
Grazing-angle studies of the MALTA2 sensor were performed both in measurements and simulations to investigate the dependence of efficiency
and cluster size on the incident angle, as well as to estimate the active depth of the depleted region of the sensor. The DUT plane was 
rotated with respect to the beam line, yielding an incident angle ($\upalpha$) of particle beam relative to the row direction of 
the pixel matrix, ranging from $0^{\circ}$ to $60^{\circ}$ in steps of $5^{\circ}$ . For each angle, 1 million events were collected 
for studies of the 
detection efficiency and the cluster size.
\begin{figure}[htb!]
  \centering
    \subfigure[]{
    \includegraphics[width=0.32\textwidth]{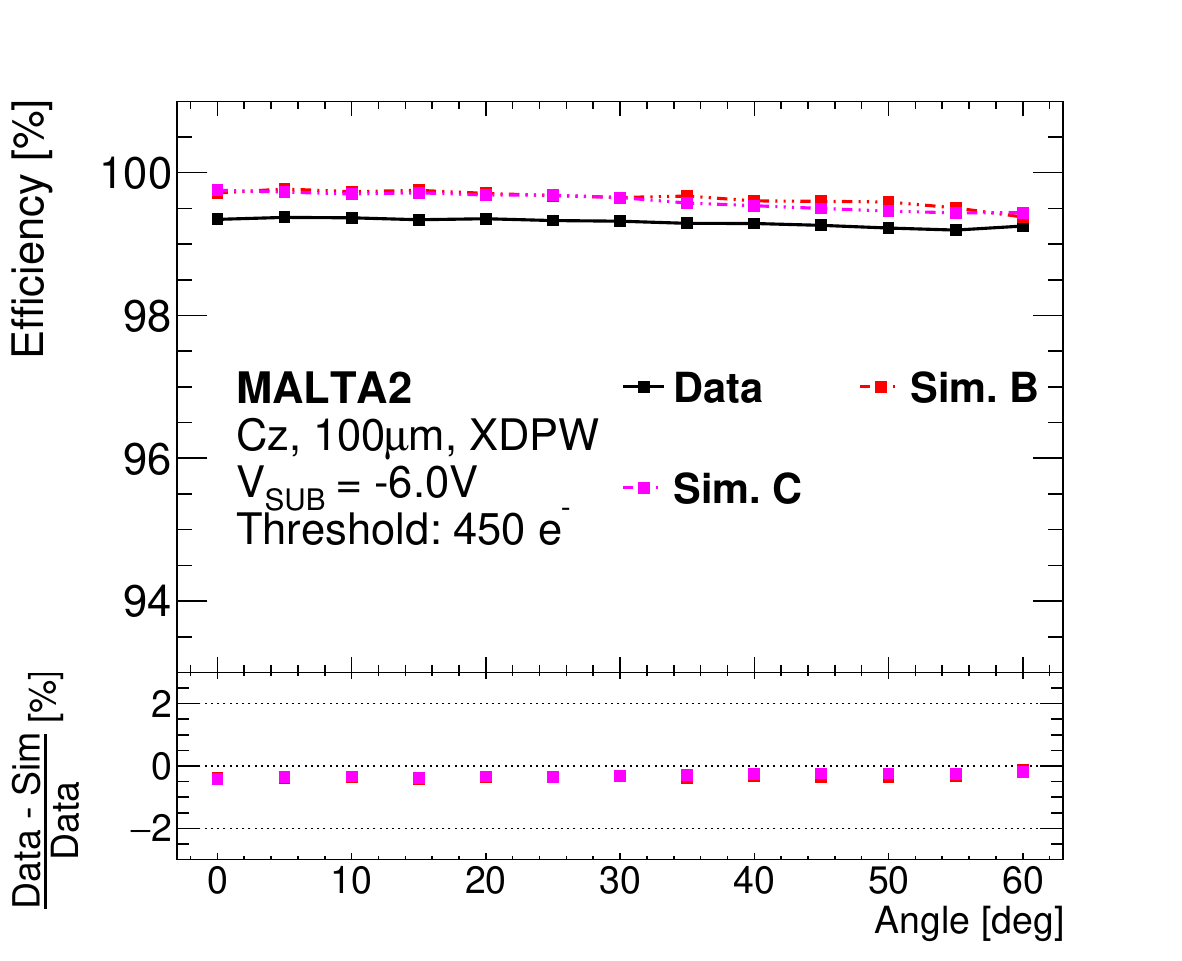}
    \label{fig:ga:eff450}
  }%
  \subfigure[]{
    \includegraphics[width=0.32\textwidth]{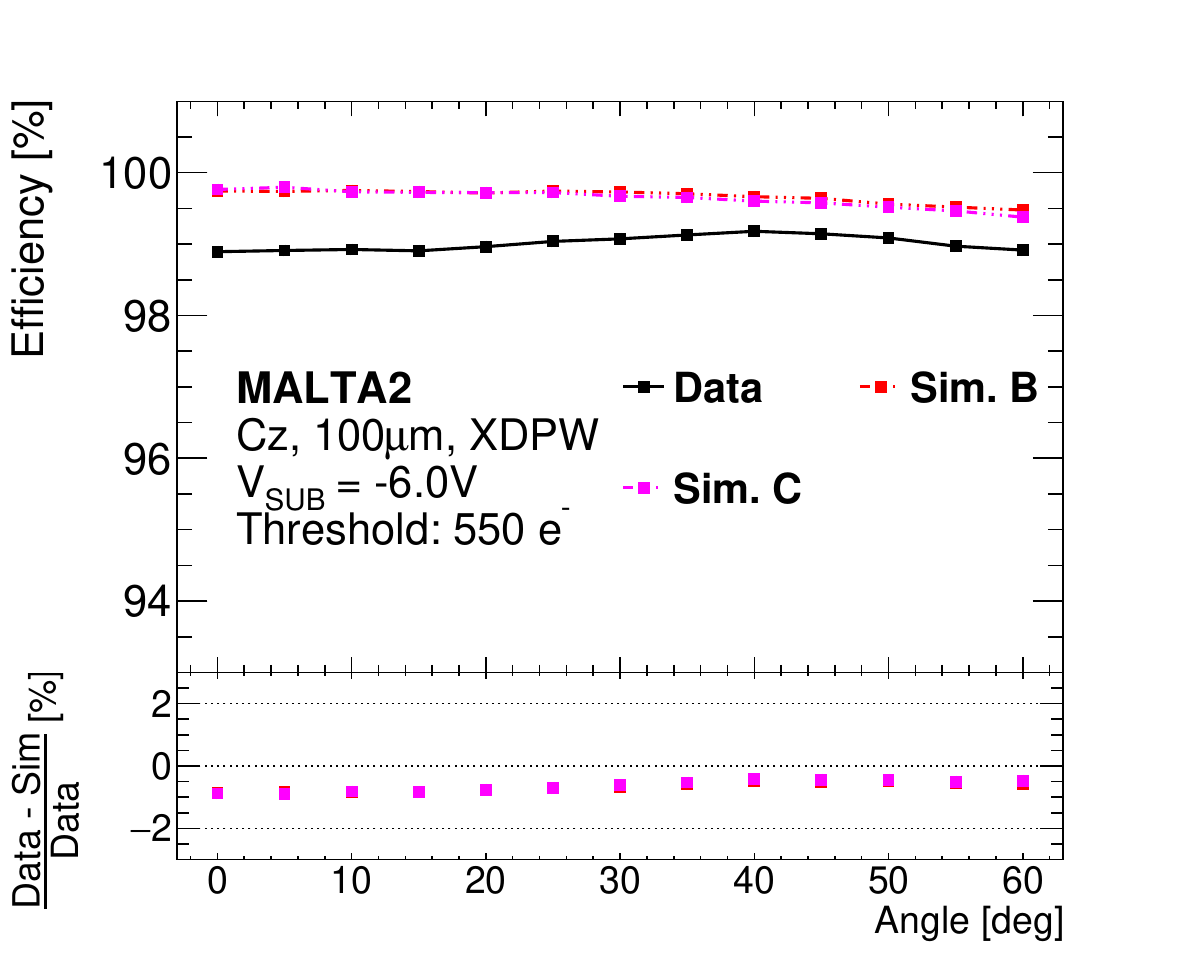}
    \label{fig:ga:eff550}
  }%
  \subfigure[]{
    \includegraphics[width=0.32\textwidth]{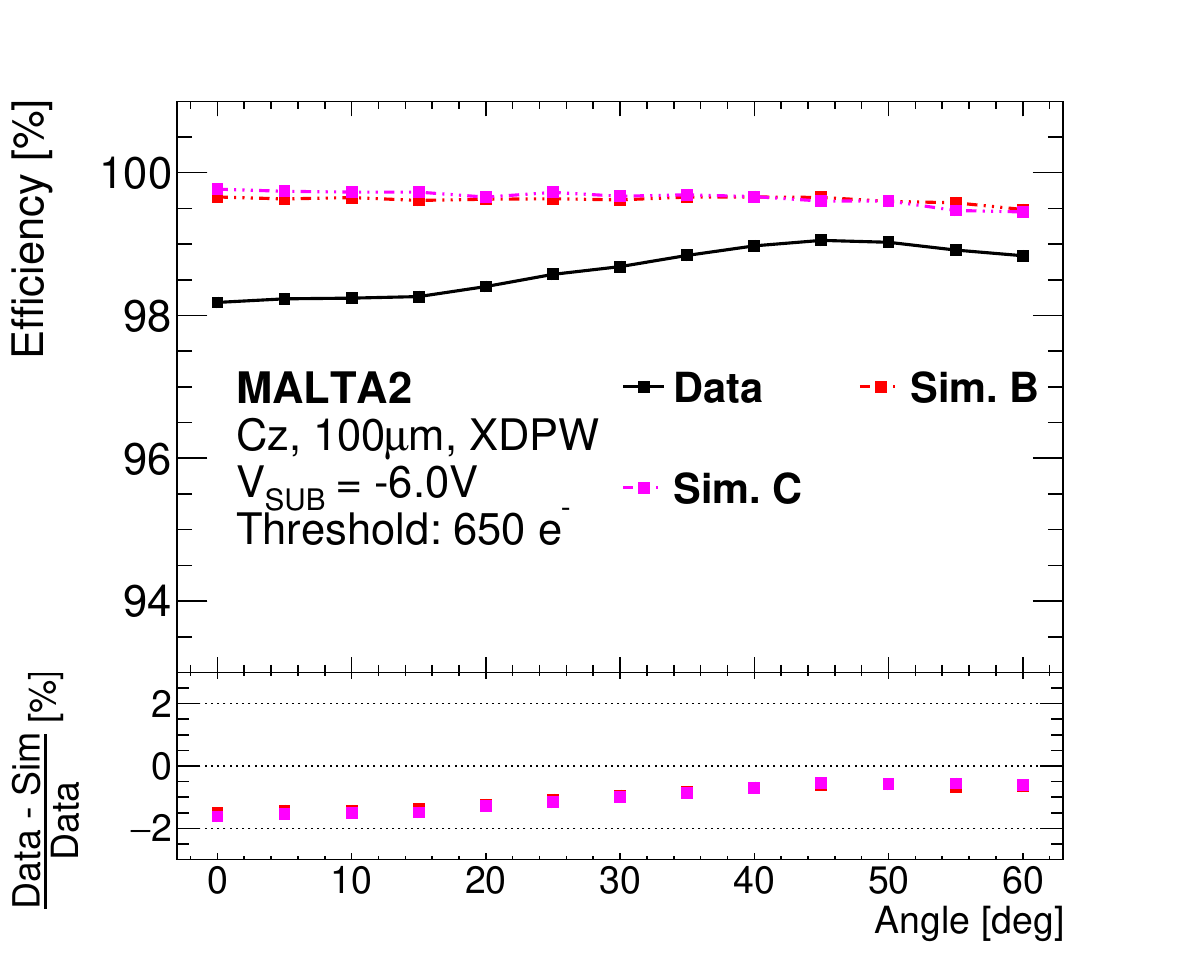}
    \label{fig:ga:eff650}
  }
  \subfigure[]{
    \includegraphics[width=0.32\textwidth]{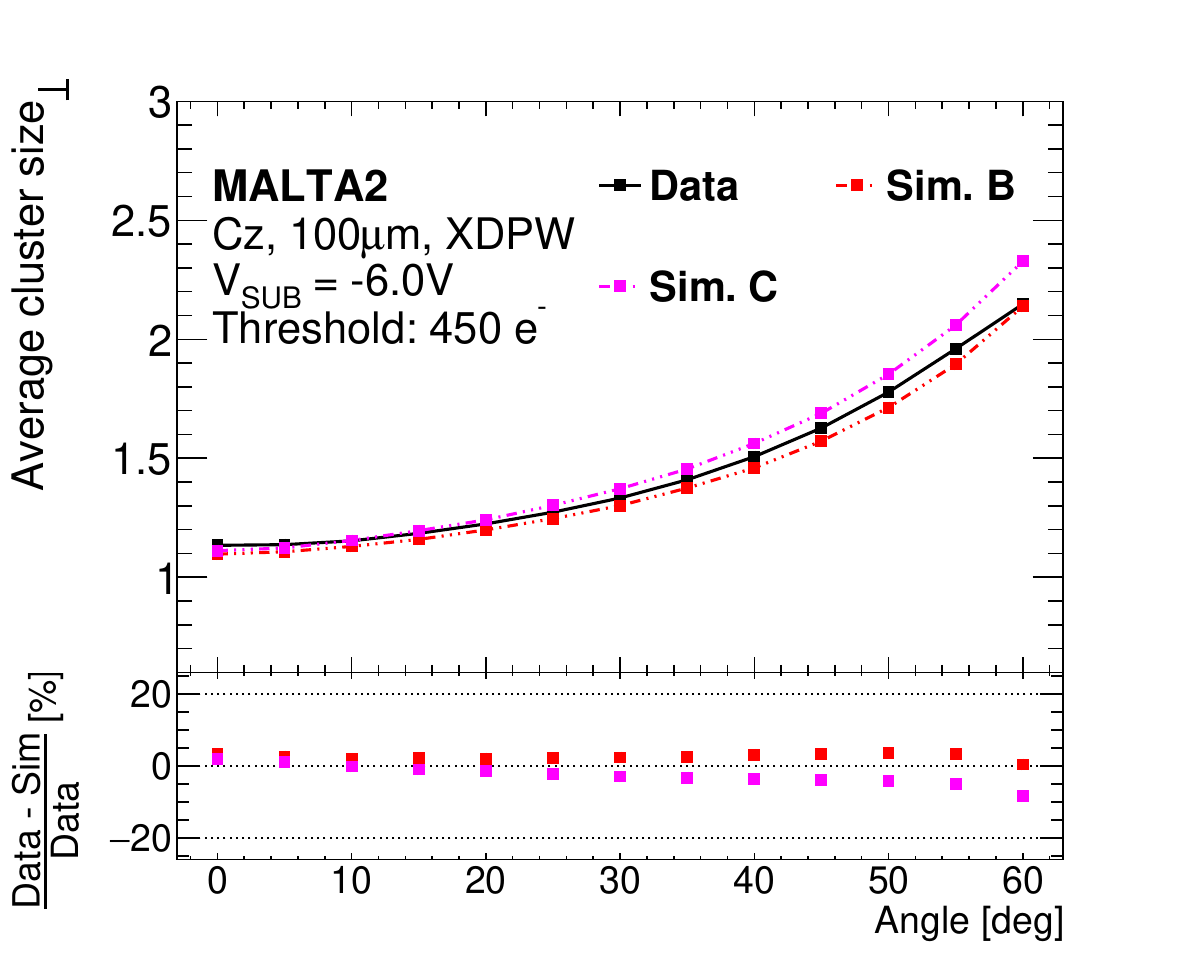}
    \label{fig:ga:clsizx450}
  }%
  \subfigure[]{
    \includegraphics[width=0.32\textwidth]{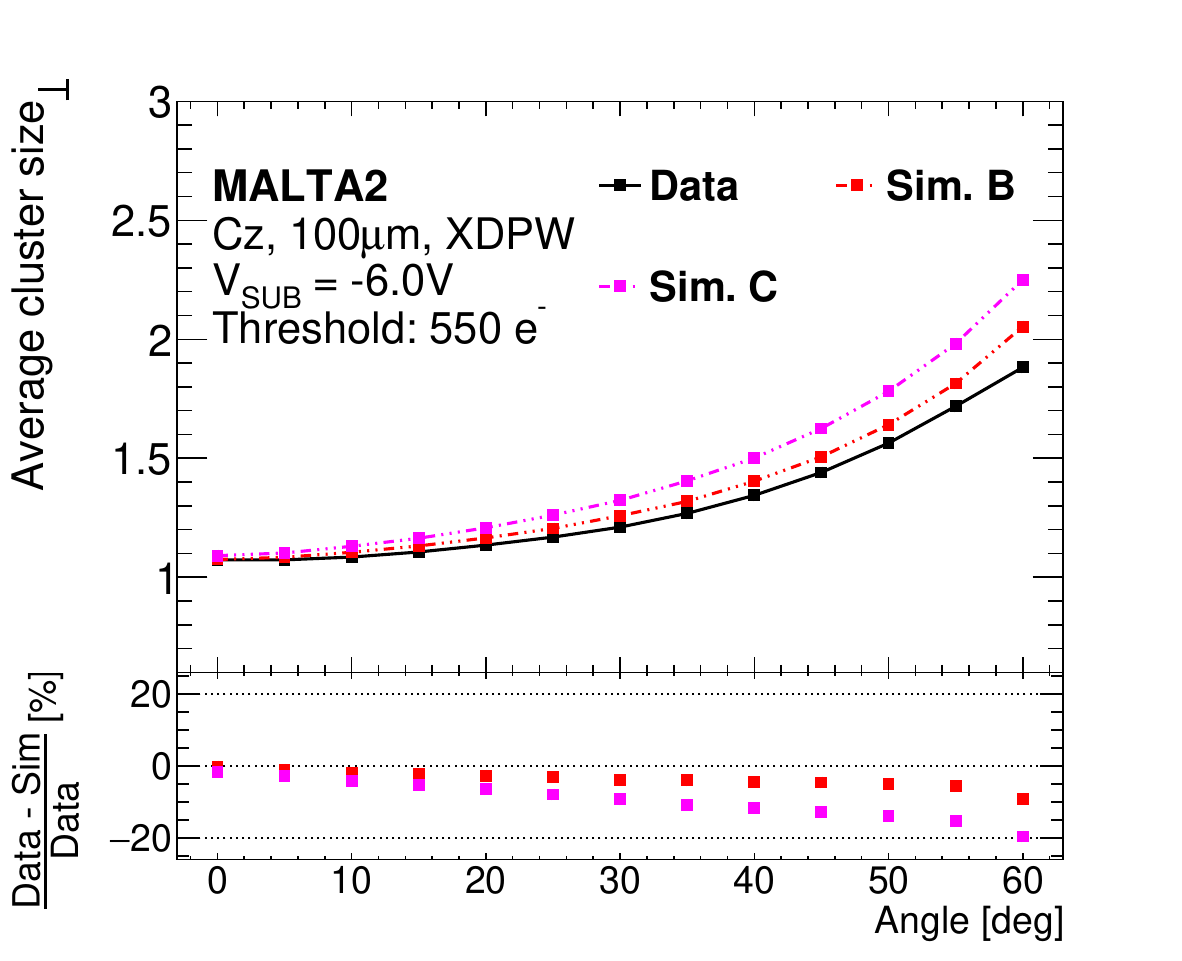}
    \label{fig:ga:clsizx550}
  }%
  \subfigure[]{
    \includegraphics[width=0.32\textwidth]{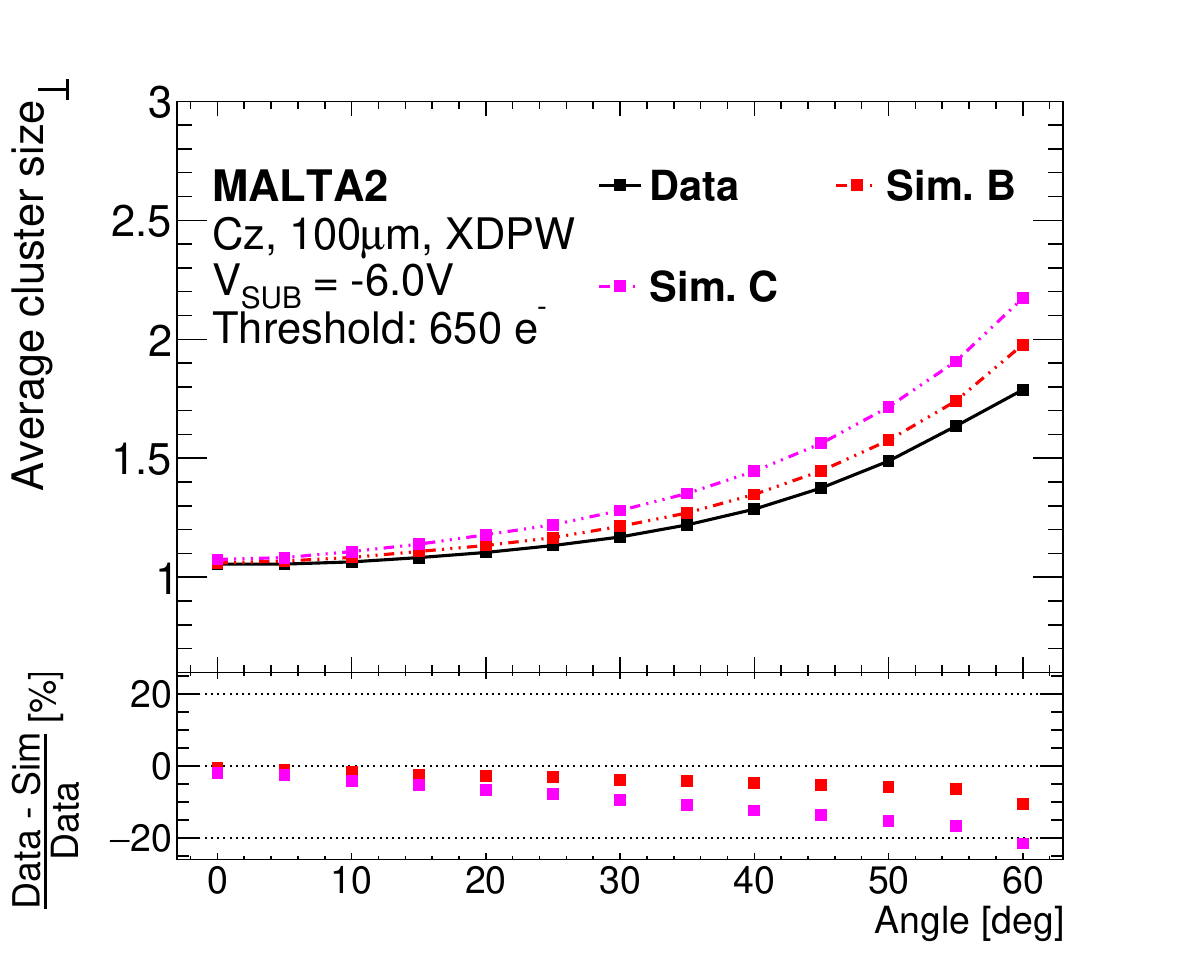}
    \label{fig:ga:clsizx650}
  }
  \caption{The detection efficiency, together with the cluster size perpendicular to the rotation axis, as a function of the
  grazing angle for measurements and simulations with N-dop concentrations B and C, at thresholds of 450\,$\text{e}^-$ (a, d), 
  550\,$\text{e}^-$  (b, e), and 650\,$\text{e}^-$ (c, f).The bias voltage is set to -6\,V. }
  \label{fig:ga}
\end{figure}
 
\fig{ga} compares the detection efficiency, combined with the cluster size ($\text{ClSize}_{\perp}$) 
perpendicular to the rotation axis, between measurements and simulations (with N-dop concentrations B and C) at thresholds of 
450\,$\text{e}^{-}$, 550\,$\text{e}^{-}$ and 650\,$\text{e}^{-}$. 
The measured and simulated detection efficiencies as a function of grazing angles, shown in  (a), (b) and (c), 
exceed 98\% and 99\% respectively, 
across all angle and threshold combinations, yielding good agreement with a relative discrepancy of less than 2\%.
At higher thresholds, 650\,$\text{e}^{-}$ for example, the measured efficiency increases slightly against the grazing angle, and 
achieves a maximum value of approximately 99.1\% at $\sim 45^{\circ}$, and then decreases as the angle increases further.
This behaviour is attributed to the interplay between enhanced charge collection and hits-merging effects at larger angles. 
On one hand, longer trajectories of incident particles at larger rotation angles contribute to more charge 
generation and collection, thereby enhancing the detection efficiency and cluster size. On the other hand, the increased cluster 
size results in more frequent hits-merging, which causes mismatches between reconstructed tracks and clusters on the DUT.
The simulated efficiency is essentially saturated ($>99.6\%$) against grazing angle, with a maximum decrease of 0.2\% at 
60$^{\circ}$ due to hits-merging.

For $\text{ClSize}_{\perp}$, the measurements and simulations show good agreement at a threshold of 450\,$\text{e}^-$ (shown in (d)), 
with relative variations below 8\% across 
all angles. However, the differences between simulations and measurements widen as the incident angle increases, and becoming 
more pronounced at higher thresholds (see (e) and (f)). N-dop 
concentration B matches the measurements better, with relative variations below 5\% across angles at 450\,$\text{e}^-$ and 
below 10\% at higher thresholds, while C shows slightly larger deviations.

\begin{figure}[htb!]
  \centering
  \includegraphics[width=0.6\textwidth, origin=c]{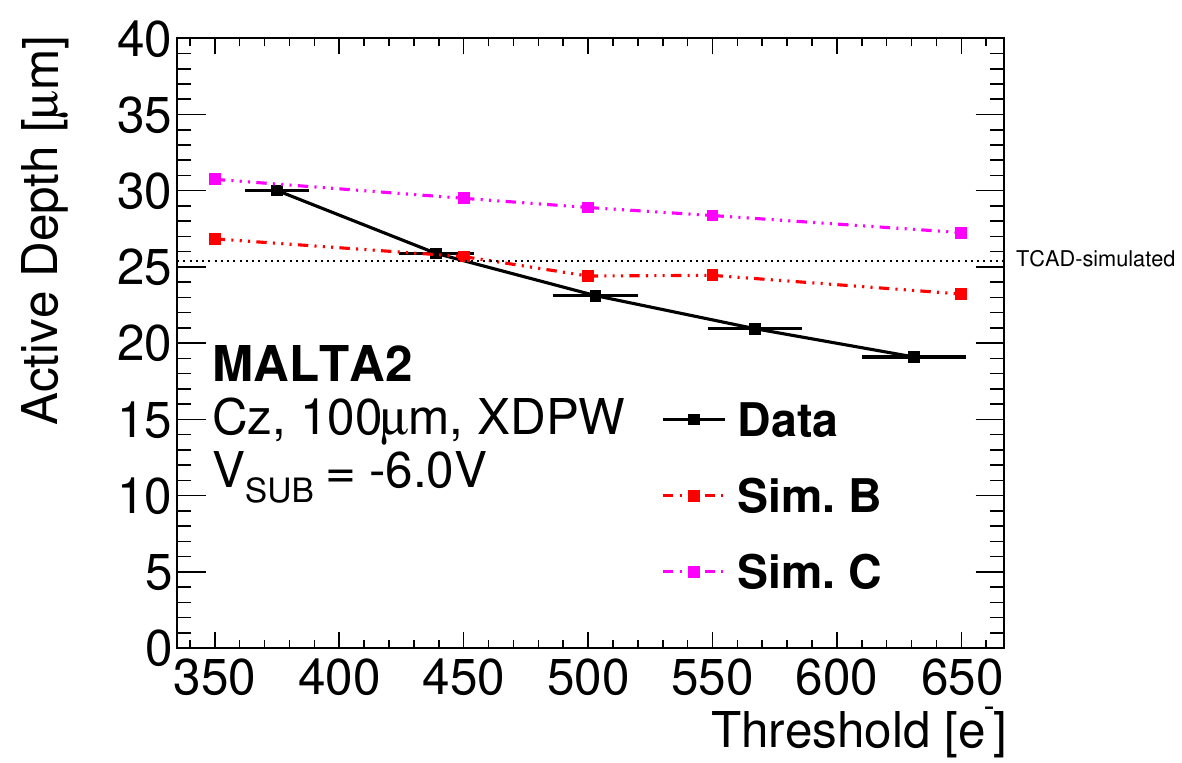}
  \caption{Estimated active depth of the depleted region of the MALTA2 sensor in measurements and simulations with N-dop concentrations B
  and C, as a function of the threshold. The bias voltage is set to -6\,V. \textsf{TCAD-simulated} represents the thickness of the depleted 
  region in TCAD simulations.}
  \label{fig:activedepth}
\end{figure}

The active depth of the depleted region of MALTA2 sensor, both in measurements and simulations, is estimated using the geometric 
relation between the active depth (\textit{D}) and $\text{ClSize}_{\perp}$, described as:
\begin{equation}
\text{ClSize}_{\perp} (\upalpha) = \frac{\text{D}}{\text{P}}\tan(\upalpha) + \text{ClSize}_{\perp}(0),
\label{eq:depth_estimation}
\end{equation}  
where \textit{P} is the pixel pitch, $\text{ClSize}_{\perp} (0)$ and $\text{ClSize}_{\perp} (\upalpha)$ represent the cluster size in 
the perpendicular direction at incident angles of 0 and $\upalpha$. For each dataset, a linear fit is performed between 
$\text{ClSize}_{\perp} (\upalpha)$ and $\tan(\upalpha)$ with $\text{ClSize}_{\perp} (0)$ fixed to the value measured. 
The active depth \textit{D} is then extracted from the slope of the fitted line. 

The estimated active depths for both measurements and simulations (with N-dop concentrations B and C), as a function of the threshold,
are shown in \fig{activedepth}. 
Again, simulations with N-dop B yield closer agreement with measurements compared those with C, with deviations within 
4\,$\upmu$m across all tested thresholds. The corresponding estimates of active-depth range from $23.2\,\pm\,0.1\,\upmu$m to 
$26.8\,\pm\,0.1\,\upmu$m, depending on the threshold, and are additionally consistent with the thickness of the depleted region 
from TCAD simulation, which is approximately 25\,$\upmu$m.

\section{Conclusion}
\label{sect:conclusion}
In this work, a simulation study of the MALTA2 sensor based on the TCAD + $\text{Allpix}^2$ framework has been presented. Using simple assumptions on 
well structures and generic doping profiles, the study began with 3D modeling and transient simulations in Sentaurus TCAD. The 
user-defined doping profiles and the resulting electric field distributions were then imported into $\text{Allpix}^2$ for high-statistics 
Monte Carlo simulations, including both DUT-only and full-telescope studies. The simulated sensor
performance -- detection efficiency, cluster size, and leading-hit ToA -- was evaluated as a function of N-dop concentration and compared 
with test-beam measurements.

TCAD simulations demonstrate efficient charge collection for both typical particle traversal scenarios 
(hit on electrode and hit at pixel corner), with total collected charges exceeding 2600\,$\text{e}^-$ and more than 70\% of the 
charge is collected within 10\,ns. The transient current responses also show a 3.5\,ns delay in the matrix configuration relative 
to the single-pixel case, confirming the finite drift time of charge carriers generated at the pixel corner.

In $\text{Allpix}^2$, doping-concentration-dependent charge collection and charge sharing are observed, arising from variations in the
electric field with N-dop concentration. Within the simulation alone, N-dop D yields the optimal charge collection and time response.
However, when compared with measurements, lower doping concentrations, specifically B and C, provide better agreement, with relative 
residuals below 25\% in detection efficiency and below 15\% in cluster size across the thresholds. Comparisons of the 1D distributions 
of cluster size and leading-hit ToA further support the methodology, while the in-pixel residual maps indicate that the descriptions of 
the doping profiles and the electric field, particularly at the pixel edges, could still be refined to achieve more realistic performance.

In the grazing-angle study, simulations with N-dop B show better agreement with measurements, especially for the cluster size perpendicular
to the rotation axis. The active depth of the depleted region, estimated from both measurements and simulations, agrees to within 4\,$\upmu$m 
across all thresholds. Simulations with N-dop B yield estimates ranging from $23.2\,\pm\,0.1\,\upmu$m to 
$26.8\,\pm\,0.1\,\upmu$m, depending on the threshold, in close agreement with the TCAD expectation of approximately 25\,$\upmu$m.

The simulations, based on basic assumptions and generic doping profiles, provide valuable insights into the sensor performance of MALTA2 
without requiring proprietary foundry information. The discrepancy between the N-dop concentration that provides the best sensor performance 
in simulation (D) and the one that best matches the measurements (B) suggests either a direction for future sensor design improvements or 
the need to include additional physical mechanisms in the simulation framework.

As the next step, additional parameters in the TCAD device model -- such as the doping concentration of the deep P‑well -- will be investigated 
to further understand the performance discrepancies observed between 
measurements and simulations. Furthermore, a radiation damage model will be introduced into the simulation chain to study the radiation‑induced 
effects on MALTA2 sensors. These extensions are expected to improve the predictive power of the simulation framework and to provide a more realistic 
description of the sensor response in the high‑radiation environment foreseen for future applications.
\section{Acknowledgements}
\label{sect:acknowledgements}
This project has received funding from the European Union’s Horizon Research and Innovation program under Grant Agreement 
numbers 101004761 (AIDAinnova), 675587 (STREAM), and 654168 (IJS, Ljubljana, Slovenia).

\newpage
\appendix
\section{$\text{Allpix}^2$ simulation configurations}
\label{app:appendix}

This Appendix contains the example configurations used in $\text{Allpix}^2$ simulations in this work.

\subsection{Sensor geometry and telescope configuration}
\label{app:appendix:sensor}

\begin{lstlisting}[caption={MALTA2 sensor configuration}, label={lst:sensor_config}]

type = "monolithic"
geometry = "pixel"
sensor_material = "silicon"
number_of_pixels =  224 512
pixel_size = 36.4um 36.4um

sensor_thickness = 100um
sensor_excess_top = 0.1mm 
sensor_excess_bottom = 0.1mm 
sensor_excess_left = 0.1mm 
sensor_excess_right = 0.1mm 

[implant]
type = "frontside"
size = 2.0um 2.0um 1.0um
\end{lstlisting}

An example configuration for MALTA2 sensor model is shown in \lst{sensor_config}. The sensor is a monolithic pixel sensor 
with $224\,\times\,512$ pixels, pitch of $36.4\,\times\,36.4~\upmu\text{m}^2$, and a thickness of 100\,$\upmu$m. 
It has an excess of 0.1\,mm on 
each side. The implant is a frontside implant with dimensions of 2.0\,$\upmu$m $\times$ 2.0\,$\upmu$m $\times$ 1.0\,$\upmu$m, 
matching the collecting-diode size of the MALTA2 sensor. \lst{telescope_config} presents the configuration of the MALTA 
telescope and the MALTA2 DUT sample, with the corresponding geometric parameters listed in \tab{telescope_geo}. 
The sensor model for the DUT is named 
\texttt{malta2\_simple}, with detailed configuration shown in \lst{sensor_config}. The sensor model for the telescope 
planes (\texttt{SPS\_P1\_W4R12} for example) is MALTA, which is similar to MALTA2 but in a larger matrix 
of $512\,\times\,512$ pixels.

\begin{lstlisting}[caption={MALTA Telescope configuration}, label={lst:telescope_config}]
[plane0]
type = "SPS_P1_W4R12"
position = 0 0 0mm
orientation = 0deg 0deg 90deg

[plane1]
type = "SPS_P2_W9R11"
position = 0 0 80mm
orientation = 0deg 0deg 90deg

[plane2]
type = "SPS_P3_W7R12"
position = 0 0 160mm
orientation = 0deg 0deg 90deg

[dut]
type = "malta2_simple"
position = 0 0 642mm
orientation = 0deg 0deg 90deg

[plane3]
type = "SPS_P4_W12R1"
position = 0 0 940mm
orientation = 0deg 0deg 90deg

[plane4]
type = "SPS_P5_W10R1"
position = 0 0 1020mm
orientation = 0deg 0deg 90deg

[plane5]
type = "SPS_P6_W7R0"
position = 0 0 1100mm
orientation = 0deg 0deg 90deg
\end{lstlisting}

\subsection{Charge carrier generation}
\label{app:appendix:generation}
To simulate the charge generation in the silicon sensor, a \texttt{DepositionGeant4} module is used. This module 
allows the use of Geant4 particles for energy deposition. The number of electron-hole pairs created by a 
given energy deposition is calculated using the mean pair creation energy \cite{ChargeCreation}, fluctuations are modeled 
using a Fano factor assuming Gaussian statistics \cite{FanoFactor}. 

Shown in \lst{charge_generation}, a 120\,GeV/c proton beam with an elliptical cross section (2.0\,mm\,$\times$\,4.4\,mm) is
generated. The beam divergence is set to 0.5\,mrad in both directions in the transverse plane. The physics list 
\texttt{FTFP\_BERT\_EMZ} \cite{G4_3} is used, where the \texttt{FTFP\_BERT} model handles high-energy hadrons 
(120\,GeV/c proton in this case), and \texttt{EMZ} enables accurate electromagnetic physics models. The parameter 
\texttt{max\_step\_length} defines the maximum step length in the Geant4 simulation; a value of 2\,$\upmu$m is chosen 
as a trade-off between accuracy and speed.

\begin{lstlisting}[caption={Geant4-based charge deposition configuration}, label={lst:charge_generation}] 
[DepositionGeant4]
physics_list = FTFP_BERT_EMZ
particle_type = "Proton"
source_energy = 120GeV
source_position = 0mm 0mm -100mm
source_type = "beam"
beam_shape = "ellipse"
beam_size = 2.0mm, 4.4mm
beam_direction = 0 0 1
number_of_particles = 1
max_step_length = 2um
beam_divergence = 0.5mrad 0.5mrad
\end{lstlisting}

\subsection{Import of electric field and doping profile from TCAD}
\label{app:appendix:mesh}
$\text{Allpix}^2$ provides a mesh converter tool, \texttt{mesh\_converter}, to convert \texttt{DF-ISE} files from TCAD simulation, 
containing doping and electrical information, into the mesh models in binary APF-format or legacy INIT-format 
(the latter is used in this work). As shown in \lst{field}, the modules \texttt{ElectricFieldReader} and \texttt{DopingProfileReader} 
are responsible for importing the electric field and doping profile in mesh models, respectively. To achieve an effective mapping of 
the field in the sensor, the keyword \texttt{PIXEL\_FULL} is provided for parameter \texttt{field\_mapping} for a single-pixel 
layout, and \texttt{PIXEL\_FULL\_INVERSE} for a $2\,\times\,2$ pixel matrix layout. The field depth is set to 100\,$\upmu$m, which 
equals the thickness of the sensor substrate. 

\begin{lstlisting}[caption={Configurations for Electric field and doping concentration}, label={lst:field}]
[ElectricFieldReader] # DUT
name = "dut"
output_plots = false
field_mapping = PIXEL_FULL_INVERSE
model = "mesh"
field_depth = 100um
file_name = "MaltaEfield_map_MTX_-6V_ElectricField.init"

[DopingProfileReader]
name = "dut"
model = "mesh"
field_mapping = PIXEL_FULL_INVERSE
doping_depth = 100um
output_plots = false
file_name = "Malta_MTX_XDPW_DopingConcentration.init"
\end{lstlisting}

\subsection{Charge carrier propagation}
\label{app:appendix:propagation}
A \texttt{GenericPropagation} module is applied for charge carrier propagation in this work, with
its configuration shown in \lst{propagation}. The \texttt{masetti\_canali} mobility model 
\cite{MasettiMobility} and the combined \texttt{Shockley-Read-Hall} and \texttt{Auger} 
\cite{AugerRecombination1, AugerRecombination2} recombination model are used to handle the 
doping-concentration-dependent charge carrier transport. The parameter 
\texttt{charge\_per\_step} defines the number of charge carriers from a single energy deposition that are propagated 
as a group. A value of 40 is chosen as a balance between precision and computational cost.

\begin{lstlisting}[caption={Configuration of charge propagation}, label={lst:propagation}]
[GenericPropagation]
temperature = 300K
mobility_model = "masetti_canali"
recombination_model = "srh_auger"
charge_per_step = 40
integration_time = 10ns
propagate_electrons = true
\end{lstlisting}

\subsection{Charge transfer}
\label{app:appendix:transfer}
The \texttt{PulseTransfer} module aggregates propagated charges into pulses at individual pixel implants using the charge
carrier arrival time. An example configuration is shown in \lst{transfer}.

\begin{lstlisting}[caption={Pulse transfer configuration}, label={lst:transfer}]
[PulseTransfer]
collect_from_implant = true
\end{lstlisting}

\subsection{Signal digitization}
\label{app:appendix:digitization}
The front-end electronics of the MALTA2 sensor are based on a voltage-sensitive amplifier (VSA) architecture. 
In a VSA, the charge-to-voltage conversion is governed by the capacitance of the collecting diode, which acts 
analogously to the feedback capacitor in a charge-sensitive amplifier (CSA). Since $\text{Allpix}^2$ does not 
provide a dedicated VSA model, the \texttt{CSADigitizer} module (in \lst{digitization}) is used as a 
behavioral proxy, with the \texttt{feedback\_capacitance} set to 5\,fF, corresponding to the MALTA2 collecting
diode capacitance \cite{MALTA2_FE}.  
\begin{lstlisting}[caption={Signal digitization configuration}, label={lst:digitization}]
[CSADigitizer]
model = "simple"
feedback_capacitance = 5e-15C/V
rise_time_constant = 22ns
feedback_time_constant = 65ns
ignore_polarity = true
\end{lstlisting}
\begin{figure}[htb!]
  \centering
    \subfigure[]{
    \includegraphics[width=0.31\textwidth]{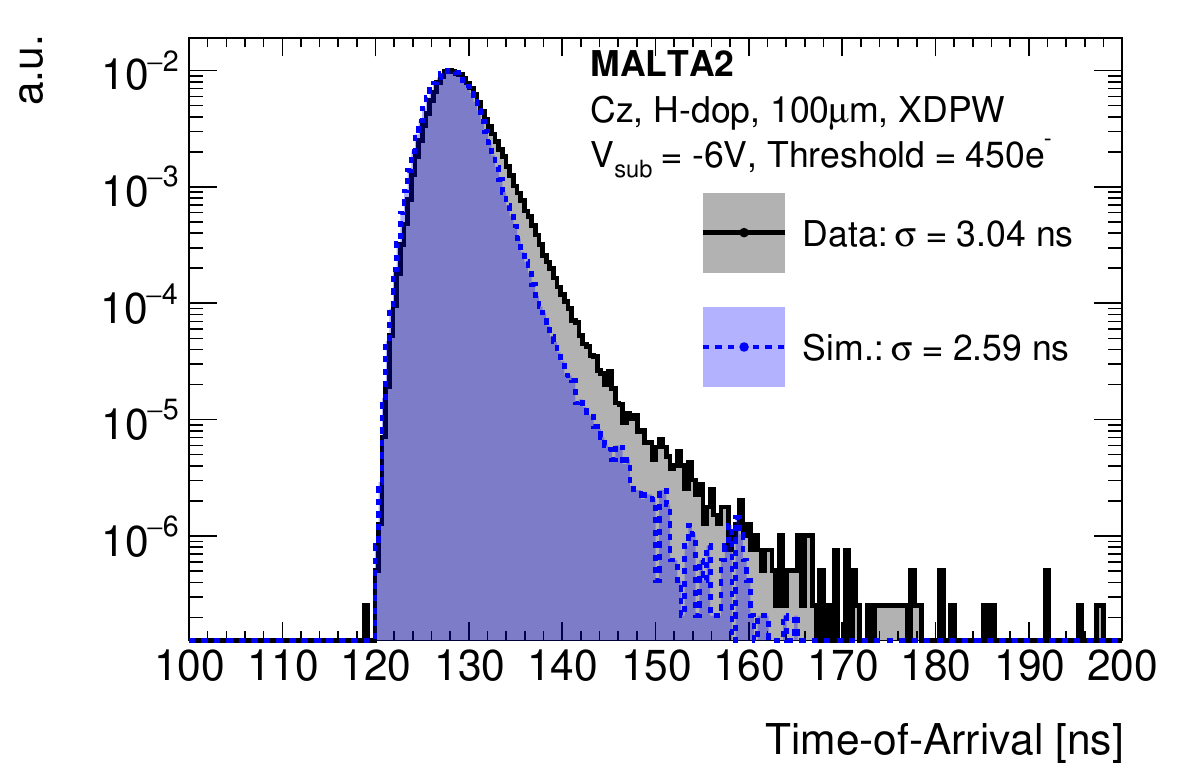}
    \label{fig:digi:toa450}
  }
    \subfigure[]{
    \includegraphics[width=0.31\textwidth]{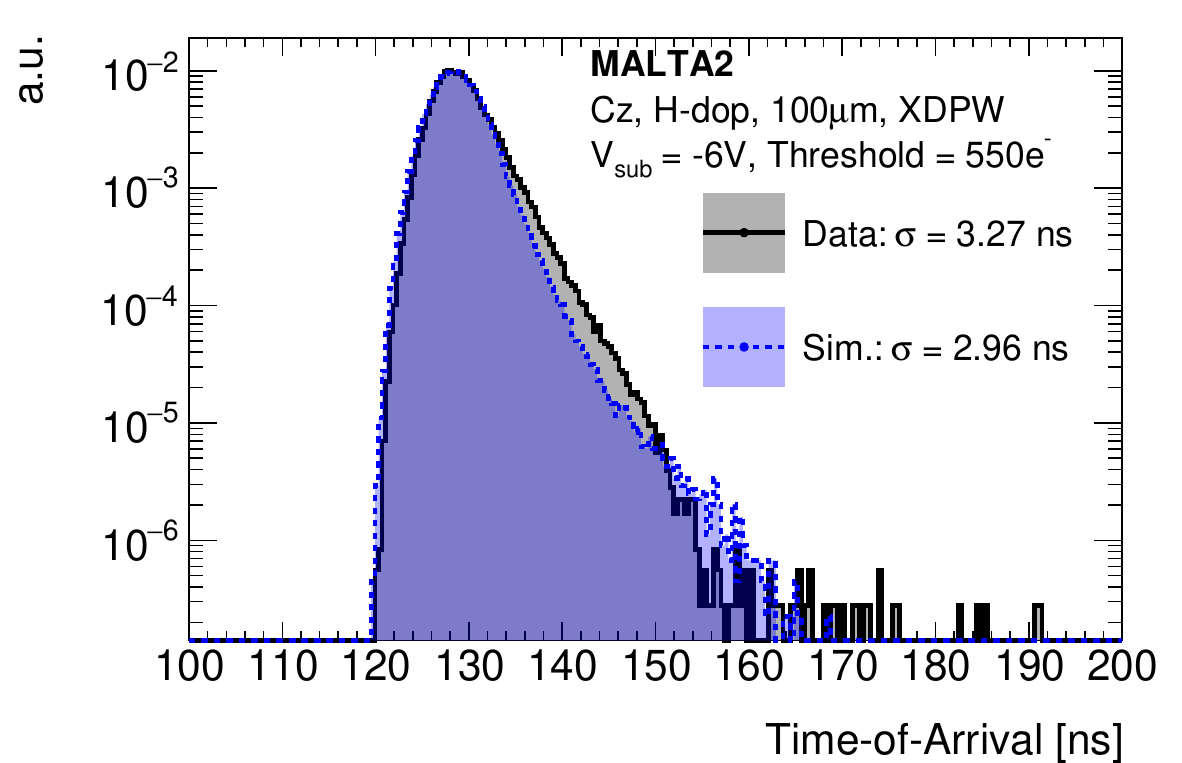}
    \label{fig:digi:toa550}
  }
  \subfigure[]{
    \includegraphics[width=0.31\textwidth]{Time-of-Arrival_comparison_thr650.pdf}
    \label{fig:digi:toa650}
  }

  \caption{Distributions of leading-hit ToA for measurements (black) and simulations (blue), at (a)
  thresholds of 450\,$\text{e}^-$, (b) 550\,$\text{e}^-$ and (c) 650\,$\text{e}^-$. The simulated 
  distributions are peak-aligned to the measurement to account for the external delay in the 
  measurement. The bias voltage is -6\,V.}
  \label{fig:digi}
\end{figure}
The parameters, \texttt{rise\_time\_constant} ($\tau_r$) and 
\texttt{feedback\_time\_constant} ($\tau_f$), define the pulse shape. They are initialized based on the work in 
Ref. \cite{MALTA2_FastSim}, and set to 22\,ns ($\tau_r$) and 65\,ns ($\tau_f$) after validations by comparing the 
simulated time-of-arrival distributions to measurements at various thresholds, as described in \fig{digi}.

\bibliographystyle{JHEP}
\bibliography{MALTA2}

\end{document}